# A periodic Energy Decomposition Analysis (pEDA) method for the Investigation of Chemical Bonding in Extended Systems

Marc Raupach and Ralf Tonner[1]

Fachbereich Chemie, Philipps-Universität Marburg, Hans-Meerwein-Straße, 35032 Marburg, Germany

The development and first applications of a new periodic energy decomposition analysis (pEDA) scheme for extended systems based on the Kohn-Sham approach to density functional theory are described. The pEDA decomposes the binding energy between two fragments (e.g. the adsorption energy of a molecule on a surface) into several well-defined terms: preparation, electrostatic and dispersion interaction, Pauli repulsion and orbital relaxation energies. The pEDA presented here for an AO-based implementation can handle restricted and unrestricted fragments for 0D to 3D systems considering periodic boundary conditions with and without the determination of fragment occupations. For the latter case, reciprocal space sampling is enabled. The new method gives comparable results to established schemes for molecular systems and shows good convergence with respect to the basis set (TZ2P), the integration accuracy and k-space sampling. Four typical bonding scenarios for surface adsorbate complexes were chosen to highlight the performance of the method representing insulating (CO on MgO(001)), metallic ($H_2$ on M(001), M = Pd, Cu) and semiconducting (CO and $C_2H_2$ on Si(001)c(4x2)) substrates. These examples cover the regimes of metallic, semiconducting and insulating substrates as well as bonding scenarios ranging from weakly interacting to covalent (shared electron and donor acceptor) bonding. The results presented lend confidence, that the pEDA will be a powerful tool for the analysis of surface-adsorbate binding in the future, enabling the transfer of concepts like ionic and covalent binding, donor-acceptor interaction, steric repulsion and others to extended systems.

---

[1] Author to whom correspondence should be addressed. Electronic mail: tonner@chemie.uni-marburg.de.

## 1 Introduction

The understanding in chemistry is most often based on heuristic concepts. The most powerful and most widely used concept is chemical bonding. Classifications such as covalent, ionic or metallic bonding are central in discussing trends in different compounds and predicting new reactivity. An in-depth understanding of the chemical bond in a system therefore paves the way to predict trends and explore new reactivity. Computational methods are thereby especially helpful to complement experimental determinations of bonding parameters like structure and electron density distribution with quantitative analysis of observable and non-observable quantities.

While the discussion of chemical bonding is a major pillar for molecular chemistry,[1] the efforts to extend the concepts toward extended systems are scarcer. Still, quantitative analysis of chemical bonding in extended systems leads to greater fundamental understanding of the systems investigated and can lead to predicting new trends and reactivity for ab initio material's design and functionalization of surfaces and interfaces.[2] The advent of quantum chemistry introduced the wave function and an intrinsically delocalized picture of the electronic structure in molecules and solids. Thus, the need to regain descriptions of localized phenomena like chemical bonding was immediately evident. Subsequently, a wide range of methods for analysing the electronic structure of compounds was developed. These can be broadly classified as follows: (i) electron-density- (real space) based and (ii) orbital- (Hilbert space) based methods, (iii) methods that model experimentally observable quantities and (iv) energy-based methods. For molecular quantum chemistry, all these methods have been thoroughly explored in the past.[3] For extended structures, the available approaches and experiences are more limited[4] and shall be briefly summarized.

Direct space methods including the Quantum Theory of Atoms in Molecules and Crystals (QTAIMAC)[5], the electron localization function (ELF)[6] and similar approaches have been utilized to solve many questions of solid-state chemistry, also owing to the fact that they can be applied to experimentally derived electron densities as well.[7] These methods have the advantage of a well-defined quantum mechanical framework but they do not provide rigorous answers e.g. about the existence of a chemical bond in a system.[8] This information can only be gained by interpretation of the results based on chemical experience and knowledge.

Hilbert space methods do not analyze the electron density as a whole, but rely on the interpretation of atomic contributions to molecular properties. To this end, they are usually based on atomic orbitals resulting from the basis set approximation in many quantum chemical methods. The most heavily used Hilbert space methods are the approaches to determine partial charges of atoms in molecules and solids[9] despite their known shortcomings.[10] They are commonly used to determine bond polarity and ionic/covalent character of bonding although charge is a scalar quantity which does not carry information about the charge distribution, which can significantly alter interpretation.[11]

A popular method in surface science is the analysis of density of states (DOS) and partial DOS (pDOS) as more easily accessible representation of the band structure.[12] Closely related, an analysis of bonding and antibonding character of orbitals in an energy interval of a DOS/pDOS can be carried out with the crystal orbital overlap population (COOP)[13] and the related COHP method.[14] Other important methods are Wannier-Type Atomic Orbitals (WTAO)[10], the d-band model by Hammer and Nørskov[15], reactivity indices[16] and concepts[17] or computation of spectroscopic properties.[18] Furthermore, natural bond orbital (NBO) analysis was recently extended to periodic systems by Dunnington et al.[19] and Galeev et al.[20] Bond order methods date back to Pauling[21] and the "mobile order" defined by Coulson[22] and were reformulated many times.[23]

None of the methods described above addresses the question of energetic contributions to chemical bonds, although energy changes are the ultimate driving forces in bond formation. The quantitative description of these contributions is far less developed. One exception is the Energy Density Analysis by Nakai et al., which relies on partitioning of the total energy into atomic contributions and has recently been extended to periodic systems.[24]

For molecular calculations, the energy-based analysis methods are well established and have been used successfully in the past.[25] The symmetry-adapted perturbation theory (SAPT)[26] method is the most popular method following a perturbational approach. Other include the method by Hayes and Stone.[27] More prominent are the variational methods employing a block-partitioning of the Fock matrix. A not exhaustive list of approaches in this category is the Constrained Space Orbital Variation (CSOV) scheme developed by Bagus and Illas,[28] the natural energy decomposition analysis (NEDA) by Glendening and Weinhold,[29] the reduced variational self-consistent field (RVS-SCF) approach by Stevens and Fink,[30] the AIM-based decomposition scheme by Francisco et al.[31], a steric analysis by Liu[32] and the schemes by Mayer[33] and Korchoviec.[34] A slightly different method employing absolutely localized molecular orbitals is the recently developed ALMO-EDA by Head-Gordon and co-workers[35] which can also be applied to correlated wave functions.[36] The block-localized wave function method (BLW-ED) based on valence-bond theory has been put forward by Mo.[37]

The most heavily used variational method is the Energy Decomposition Analysis (EDA)[38] based on developments by Morokuma[39] and the related extended transition state (ETS) method by Ziegler and Rauk[40] (jointly called EDA in the following). Variations of this method are found in the recent literature.[41] A valuable extension is the recently developed EDA-NOCV (Natural Orbitals for Chemical Valence) method, which offers additional insight by providing charge and energy analysis in a combined fashion.[42] Charge transfer and charge redistribution has even been used in the past to define the formation of a surface chemical bond upon adsorption but without the ability for quantitative analysis.[4c] The strength of these methods is the ability to quantify different bonding contributions in terms of energy. Only the CSOV and the EDA were extended to periodic systems but were only used to a limited extent in cluster-based[43] and pilot studies.[44] Thus until today, quantitative energy-based analysis of chemical bonding in periodic systems has neither been documented for surface-adsorbate interactions in a broader study.

We now introduce the periodic Energy Decomposition Analysis (pEDA) method which decomposes the interaction energy in periodic systems into chemically intuitive terms. Theory and implementation of the method are outlined, the validity and accuracy is tested and the method is applied to prototype surface science questions spanning different bonding scenarios.

## 2 Theoretical Background and Implementation

### 2.1 The EDA method

First, we outline the working principle of the molecular EDA method as developed by Ziegler/Rauk[40] and Morokuma[39]. The approach used in the EDA is the investigation of the intrinsic bond energy for the interaction of two fragments A and B forming a molecule AB by separating the bond formation process into several sub-steps. The bond dissociation energy $\Delta E_{bond}$ (which is the negative of the dissociation energy without zero-point vibrational corrections $D_e$) is thereby composed of the energy changes from a promotion step and an interaction step (Scheme 1a) with the respective energy terms ($\Delta E_{prep} + \Delta E_{int}$).

$$\Delta E_{bond} = \Delta E_{prep} + \Delta E_{int} \qquad (1)$$

Scheme 1

In the initial step, the ground state (GS) configuration of the fragments $A^{GS}$ and $B^{GS}$ are distorted to the particular reference state A and B they have in the molecule AB. This includes geometric distortion and electronic excitation and requires a preparation energy $\Delta E_{prep}$ which is given by the energies of relaxed and distorted fragments.

$$\Delta E_{prep} = \left(E_A^{GS} + E_B^{GS}\right) - \left(E_A + E_B\right) \qquad (2)$$

The promoted fragments can be represented by Slater determinants $\Psi_A$ and $\Psi_B$ built from fragment orbitals at A and B respectively. They are now interacting to form AB. In this step, the intrinsic bond energy $\Delta E_{int}$ results from the energy difference between the molecule ($E_{AB}$) and the prepared fragments ($E_A$, $E_B$). The basic idea of the EDA scheme now is the partitioning of the interaction energy $\Delta E_{int}$ into well-defined terms (Scheme 1a).

$$\begin{aligned} \Delta E_{int} &= E_{AB} - (E_A + E_B) \\ &= \Delta E_{elstat} + \Delta E_{Pauli} + \Delta E_{orb} \end{aligned} \qquad (3)$$

First, the distorted fragments are brought from infinity into the position they occupy in the molecule AB without optimisation of the resulting wave function. The associated energy change for this step is the quasiclassical electrostatic interaction ($\Delta E_{elstat}$) between the two charge distributions $\rho_A$ and $\rho_B$ of both fragments. This term is usually attractive and gives a quantitative estimate of the electrostatic bonding contributions which are neglected in a purely orbital-interaction based analysis.[45] The energy after this step corresponds to a simple product wave function $\{\Psi_A \Psi_B\}$ which does not fulfil the Pauli Exclusion Principle. In the next step, this product wave function is antisymmetrized ($\hat{A}$) and normalized ($N$). This leads to the intermediate wave function $\Psi^0$ with the energy $E^0$.

$$\Psi^0 = N\hat{A}\{\Psi_A \Psi_B\} \qquad (4)$$

The orbitals making up $\Psi^0$ are obtained via separate Löwdin orthogonalization of occupied and virtual fragment orbitals followed by Gram-Schmidt orthogonalization of occupied on virtual orbitals. The corresponding term $\Delta E_{Pauli}$ is destabilizing since the orthogonalization procedure adds additional nodes to the orbitals and thus leads to an increase of the kinetic energy.[46] In the final step, the frozen molecular orbitals of the intermediate wave function are allowed to relax and the optimal wave function $\Psi_{AB}$ for the molecule AB with the energy $E_{AB}$ is found. This energy ($\Delta E_{orb}$) can be expressed in terms of the density $\rho^0$ of the intermediate wave function $\Psi^0$ and the density $\rho_{AB}$ of the final wave function $\Psi_{AB}$.

$$\Delta E_{orb} = E\left[\rho_{AB}\right] - E\left[\rho^0\right] \qquad (5)$$

## 2.2 Implementation of the pEDA and inclusion of a dispersion term

Following the discussion in recent years about the importance of dispersion interactions in DFT-based periodic and non-periodic computations,[47] we chose to include a dispersion term in our pEDA method. The validity of treating this term separately from the EDA has been shown before[48] and an improved performance was found.[49] Therefore we chose to obtain $\Delta E_{disp}$ via the semi-empirical correction scheme DFT-D3 put forward by Grimme et al.[56] via the difference of fragment and complex dispersion energies:

$$\Delta E_{disp} = E_{disp,AB} - (E_{disp,A} + E_{disp,B}) \tag{6}$$

The alternative approach to use dispersion-corrected density functionals (e.g. VV10[50]) would disable the separate discussion of this bonding contribution.

For the simulation of extended systems, we chose an approach based on atom centered, atomic orbital (AO) basis functions, $\{\phi_\mu\}$, to generate the corresponding Bloch functions, $\{\Phi_\mu(\mathbf{k})\}$. Linear combination of these basis functions leads to a set of crystal orbitals (CO) $\{\varphi_i(\mathbf{k})\}$ for every point in reciprocal space k, where the coefficients $c_{i,\mu}(\mathbf{k})$ are the transformation matrix elements of $\mathbb{T}(\mathbf{k})$.

$$\varphi_i(\mathbf{k}) = \sum_\mu c_{i,\mu}(\mathbf{k})\Phi_\mu(\mathbf{k}) \text{ or } \{\varphi_i(\mathbf{k})\} = \mathbb{T}(\mathbf{k})\{\Phi_\mu(\mathbf{k})\} \tag{7}$$

As a consequence of the Bloch theorem, our wave functions are now dependent on the reciprocal wave vector k. The first Brillouin zone in reciprocal space contains an infinite number of k-points but it is enough to evaluate only a sub-set of k-points to derive at a good approximation of the wave function and the density of the extended system (k-space sampling).

For every k-point the Kohn-Sham equations have to be solved to derive $\Psi_A(\mathbf{k})$ and $\Psi_B(\mathbf{k})$ which are described in the basis of the Bloch functions by matrices $\mathbb{T}_A(\mathbf{k})$ and $\mathbb{T}_B(\mathbf{k})$. The total charge density of the fragments, $\rho_A$ and $\rho_B$, is then the weighted sum of all densities at the k-points sampled:

$$\rho_A = \sum_\mathbf{k} \omega(\mathbf{k})\rho_A(\mathbf{k}) = \sum_\mathbf{k} \omega(\mathbf{k}) \sum_{spin}^{\alpha,\beta} \sum_i^{occ} \varphi_{i,A}(\mathbf{k})\varphi_{i,A}(\mathbf{k})$$

$$\rho_B = \sum_\mathbf{k} \omega(\mathbf{k})\rho_B(\mathbf{k}) = \sum_\mathbf{k} \omega(\mathbf{k}) \sum_{spin}^{\alpha,\beta} \sum_i^{occ} \varphi_{i,B}(\mathbf{k})\varphi_{i,B}(\mathbf{k}) \tag{8}$$

These charge densities are now used to calculate the electrostatic energy term, $\Delta E_{elstat}$:

$$\Delta E_{elstat} = \sum_{\nu \varepsilon A} \sum_{\mu \varepsilon B} \frac{Z_\nu Z_\mu}{|R_\nu - R_\mu|} - \sum_{\nu \varepsilon A} \int \frac{Z_\nu \rho_B(r_i)}{|R_\nu - r_i|} dr_i$$
$$- \sum_{\mu \varepsilon B} \int \frac{Z_\mu \rho_B(r_i)}{|R_\mu - r_i|} dr_i + \int \int \frac{\rho_A(r_i) \rho_B(r_j)}{|r_i - r_j|} dr_i dr_j \quad (9)$$

For the calculation of the energy terms $\Delta E_{Pauli}$ and $\Delta E_{orb}$ the definition of the orthogonalized wave functions, $\Psi^0(\mathbf{k})$, is necessary. Based on work by Philipsen and Baerends,[44a] the algorithm to construct the intermediate wave functions can be separated in three steps. The initial step is the combination of the separated eigenstates, $\Psi_A(\mathbf{k})$ and $\Psi_B(\mathbf{k})$, to form the normalized eigenstate $N\{\Psi_A(\mathbf{k})\Psi_B(\mathbf{k})\}$, which is described by expansion of the Bloch basis $\{\Phi_\mu(\mathbf{k})\}$ with the transformation matrix $\mathbb{T}_1(\mathbf{k})$ for every k-point. This transformation matrix can be further separated in two parts, which characterize the occupied COs, denoted as $\mathbb{T}_{o,1}(\mathbf{k})$, and the virtual COs, denoted as $\mathbb{T}_{v,1}(\mathbf{k})$. Now the occupied COs of A and B are orthogonalized via the Löwdin formalism. Only $\mathbb{S}_{o/o,1}(\mathbf{k})$, the overlap matrix for the occupied COs of the initial eigenstate, is needed to calculate the transformation matrix, $\mathbb{T}_{o,2}(\mathbf{k})$, for the orthogonalized, occupied fragment COs:

$$\mathbb{T}_{o,2}(\mathbf{k}) = \mathbb{T}_{o,1}(\mathbf{k}) \{\mathbb{S}_{o/o,1}(\mathbf{k})\}^{-1/2} \quad (10)$$

Then the virtual COs of A and B are orthogonalized onto the newly formed occupied fragment COs via the Gram-Schmidt formalism. Here, the overlap matrix for the orthogonalized occupied COs, $\mathbb{S}_{o/o,2}(\mathbf{k})$, is the identity matrix and the correction matrix $\mathbb{T}^{GS}(\mathbf{k})$ is described solely by the overlap matrix between occupied and virtual crystal orbitals $\mathbb{S}_{o/v,2}(\mathbf{k})$

$$\mathbb{T}_{v,2}(\mathbf{k}) = \mathbb{T}_{v,1}(\mathbf{k}) + \mathbb{T}_{o,2}(\mathbf{k})\mathbb{T}^{GS}(\mathbf{k}) \quad (11)$$

$$\text{with } \mathbb{T}^{GS}(\mathbf{k}) = \{\mathbb{S}_{o/o,2}(\mathbf{k})\}^{-1} \mathbb{S}_{o/v,1}(\mathbf{k}) = \mathbb{S}_{o/v,1}(\mathbf{k}) \quad (12)$$

The last step is the Löwdin orthogonalization of the newly formed virtual CO set $\mathbb{T}_{v,2}(\mathbf{k})$. Here $\mathbb{S}_{v/v,2}(\mathbf{k})$ is the overlap matrix for these virtual COs.

$$\mathbb{T}_{v,3}(\mathbf{k}) = \mathbb{T}_{v,2}(\mathbf{k}) \{\mathbb{S}_{v/v,2}(\mathbf{k})\}^{-1/2} \quad (13)$$

So, the occupied and the virtual space of the transformation matrix $\mathbb{T}^0(\mathbf{k})$ is constructed by $\mathbb{T}_{o,2}(\mathbf{k})$ and $\mathbb{T}_{v,3}(\mathbf{k})$ and expands the Bloch basis $\{\Phi_\mu(\mathbf{k})\}$ in terms of the orthogonalized fragment COs $\{\lambda_i(\mathbf{k})\}$.

$$\{\lambda_i(\mathbf{k})\} = \mathbb{T}^0(\mathbf{k}) \{\Phi_\mu(\mathbf{k})\} \quad (14)$$

Now the intermediate wave function, $\Psi^0(\mathbf{k})$, is described by orthogonalized fragment COs and forms the intermediate crystal density $\rho^0$, which can be assigned the energy value $E^0$.
The energy difference between the separated fragments and this intermediate eigenstate is calculated as follows.

$$\Delta E^0 = E\left[\rho^0\right] - \left(E\left[\rho_A\right] + E\left[\rho_B\right]\right) \quad (15)$$

The energy difference $\Delta E^0$ can be expressed as the differences of the kinetic energy $T$ of the electrons, the potential energy (Coulomb energy $V_{Coul}$) and the exchange-correlation energy ($V_{XC}$) between the two fragment states A and B and the orthogonalized, intermediate state $\Psi^0$.

$$\begin{aligned}\Delta E^0 &= T\left[\rho^0\right] - \left(T\left[\rho_A\right] + T\left[\rho_B\right]\right) + V_{Coul}\left[\rho^0\right] - \left(V_{Coul}\left[\rho_A\right] + V_{Coul}\left[\rho_B\right]\right) \\ &\qquad + V_{XC}\left[\rho^0\right] - \left(V_{XC}\left[\rho_A\right] + V_{XC}\left[\rho_B\right]\right) \\ &= \Delta T^0 + \Delta V^0_{Coul} + \Delta V^0_{XC}\end{aligned} \quad (16)$$

Since the resulting wave functions $\Psi^0(\mathbf{k})$ incorporates the quasiclassical, electrostatic interaction and the Pauli exclusion principle, the term $\Delta V^0_{Coul}$ comprises the Coulomb interaction due to the overlapping fragment densities, $\Delta E_{elstat}$, and the change of the Coulomb interaction due to the orthogonalization of the fragment wave functions, $\Delta V_{P,Coul}$.

$$\Delta V^0_{Coul} = \Delta V_{P,Coul} + \Delta E_{elstat} \quad (17)$$

$$\text{with } \Delta V_{P,Coul} = V^0_{Coul} - V^{\{\Psi_A \Psi_B\}}_{Coul} \quad (18)$$

Now the energy term, $\Delta E_{Pauli}$, corresponding to the Pauli exclusion principle, is calculated only by terms arising due to the orthogonalization scheme.

$$\Delta E_{Pauli} = \Delta T^0 - \Delta V_{P,Coul} - \Delta V^0_{XC} \quad (19)$$

In the last step, the intermediate wave function is allowed to relax to the final wave function $\Psi_{AB}$ with the electron density $\rho_{AB}$. The corresponding energy change is called the orbital relaxation energy term:

$$\Delta E_{orb} = E\left[\rho_{AB}\right] - E\left[\rho^0\right] \quad (20)$$

The overall interaction and bond dissociation energies can now be calculated according to eq. (1) and eq. 3 (see Scheme 1b). Occupation numbers can be set for the fragments chosen if k-space sampling is restricted to the Γ-point. The same restriction currently applies to the decomposition into spin polarized fragemtns.

## 3 Computational methodology

### 3.1 Molecular systems

Unconstrained structural optimizations for the molecular systems investigated here were carried out in the framework of DFT by employing the BP86[51] exchange-correlation functional together with the TZ2P basis set[52] applying the frozen core approximation, an accuracy setting of 5 and incorporating relativistic effects within the zeroth order regular approximation (ZORA)[53] without periodic boundary conditions (PBC). All calculations utilizing the frozen core approximation employed a "small" frozen core and are indexed by a suffix (fc) at the basis set level (e.g. TZ2P(fc)). These structures were analyzed with the EDA method as outlined above. Other basis sets (DZ, TZP, QZ4P)[52] were used for convergence studies. All non-periodic calculations (optimizations and EDA) were carried out with the ADF molecular modeling suite (version 2012).[54] Subsequently, the resulting structures were re-optimized under one-dimensional PBC with a unit cell length of 50 Å. Based on these structures, pEDA calculations were performed with k-space sampling restricted to the Γ-point.

### 3.2 Extended systems

The surface-adsorbate complexes investigated here were calculated applying two-dimensional PBC in the surface plane. For the adsorption of $H_2$ on Cu(001) and Pd(001) surfaces, the metal was approximated with two layers of atoms and a (2x3) super cell with lattice constants a(Cu) = 3.61 Å and a(Pd) = 3.95 Å. The $H_2$ molecule was placed parallel to the surface in bridging position. This corresponds to the setup used in the previous EDA study on this system.[44a]

For the adsorption of CO on the MgO(001) surface, the insulator surface was approximated with three layers of Mg atoms and the corresponding O atom layers employing a lattice parameter of 4.25 Å. Surface rippling was approximated by displacing the positions of the Mg/O atoms inward/outward by 0.001 Å, respectively. The CO molecule was placed perpendicular to the surface above an Mg atom with a Mg-C distance of 2.61 Å. Three different super cells were chosen, resulting in coverages of θ = 0.5, 0.25 and 0.125. This corresponds to the setup used in the previous CSOV study on this system.[44b]

The optimal adsorption structures of CO and $C_2H_2$ molecules on the reconstructed Si(001)c(4x2) surface were found by carrying out spin-polarized, constrained structural optimizations (applying three-dimensional PBC) with the two bottom layers of the six-layer silicon slab kept fixed. These calculations were performed using the PBE[55] functional considering dispersion effects via the DFT-D3 scheme with an improved damping function[56] (in the following PBE-D3). The projector-augmented wave (PAW) method was employed allowing for a kinetic energy cutoff of 350 eV for the plane wave basis set.[57] The Brillouin zone was sampled by a 4x2x1 Γ-centered Monkhorst-Pack k-mesh.[58] Based on the resulting structures, a $Si_{15}H_{16}$ cluster was derived, approximating the infinite surface in a finite model. While the Si positions were kept frozen, the dangling bonds were saturated by H atoms. These capping atoms were placed along the broken Si-Si bond and set to a typical Si-H bond lengths of 1.480 Å.[59]

Bonding analysis for these extended structures was carried out via pEDA calculations. If not otherwise noted, the pEDA calculations were done by employing the BP86 or PBE exchange-correlation functional with a TZ2P(fc) basis set, an accuracy setting of 5, applying ZORA and taking into account the Γ-point only.

Structural optimizations and PBE-D3 calculations of extended systems were carried out with the VASP code (version 5.2.12).[60] Bond analysis calculations employing PBC (pEDA) was carried out with a development version of the BAND code.[61]

## 4  Results

### 4.1  Bonding scenarios chosen

The pEDA was tested for a wide range of bonding scenarios by choosing three molecular and four extended test systems. For the molecular systems, three different prototypical bonding types were investigated: (i) Main group donor-acceptor bonding ($H_3N$-$BH_3$, Scheme 2a), (ii) transition metal donor-acceptor bonding (CO-$Cr(CO)_5$, Scheme 2b) and (iii) shared electron bonding ($H_3C$-$CH_3$, Scheme 2c). The fragmentation applied in the EDA and pEDA decompositions is indicated in Scheme 2. The inclusion of non-periodic systems in this convergence study enables us to validate the numerical results against the established EDA procedure and quantify the outcome of using unrestricted fragments (Scheme 3).

Scheme 2

Scheme 3

Nevertheless, the main aim of the new method is thus to investigate the chemical bonding in periodic systems. Thus we chose four different bonding scenarios: (i) Adsorption of $H_2$ on the Cu(001) and Pd(001) surface (Figure 1), (ii) adsorption of CO on the MgO(001) surface (Figure 2) and (iii) adsorption of CO (Figure 3) and $C_2H_2$ (Figure 4) on the Si(001)c(4x2) surface exhibiting the well-known buckled dimer reconstruction. On the one hand, these examples cover the regimes of conducting (i), insulating (ii) and semiconducting (iii) surfaces. On the other hand, the two examples for the semiconducting surface are typical examples for donor-acceptor (CO) and shared-electron ($C_2H_2$) binding to the surface as we will see later on (Scheme 4). The metallic system was chosen since it enables us to compare the results to the previous implementation of a periodic EDA by Philippsen and Baerends.[35a] The insulating system was investigated by periodic CSOV analysis before and thus lends the opportunity to compare these two methods.[38]

Scheme 4

Figure 1

Figure 2

Figure 3

Figure 4

## 4.2 Convergence behavior of pEDA terms

A robust analysis method needs to provide reliable results with respect to the computational parameters in the calculation. Therefore, we checked the convergence behavior of the pEDA results for the following settings: (i) basis set and frozen core usage, (ii) accuracy setting and (iii) k-space sampling.

We tested basis sets of double zeta (DZ), triple zeta (TZP, TZ2P) and quadruple zeta (QZ4P) quality with and without frozen core approximation (large and small frozen core). The general precision parameter (accuracy) determines the generation of integration points and many other parameters related to the accuracy of the results.[62] The parameter for the k-space sampling can be set to include the Γ-point only (setting 1), use a linear tetrahedron method (setting 2, 4,…) or a quadratic tetrahedron method (setting 3, 5, …).[63] The influence of these parameters will be checked in the following sections for molecular and extended systems.

### 4.2.1 Molecular systems

For the three molecular test systems the results for the convergence w.r.t. basis set size and frozen core approximation are found in Table 1. The influence of the accuracy parameter on the pEDA terms is shown in Table 2. The latter parameter turns out to be well converged with a setting of 5 in comparison to precise computations (accuracy 7) for all systems. The smaller setting (accuracy 3) leads to rather large deviations of the terms especially for the transition metal complex. Note, that the default setting is quite low (accuracy 3.5) and should thus be adjusted for pEDA analyses.

Regarding the basis set, we need to discuss two effects: The influence of the frozen core approximation (small cores in all cases, large core calculations lead to too large errors) and the size of the basis set. As was found for molecular systems in non-periodic calculations before[52] the frozen core approximation introduces only small errors in comparison to the all-electron results. For the TZ2P basis set, these errors are ≤ 0.5 %. At the same time SCF convergence is improved and the computing time is reduced by approximately 30%. For the double-ζ basis set without polarization functions (DZ), the deviations are much larger – notably for the transition metal complex. Here, the frozen core region is too large for the small and inflexible DZ basis set.

Table 1

Table 2

The results for the basis set convergence require a closer look. The DZ basis set is in all systems not sufficient to deliver converged pEDA energy terms and large errors remain also due to the missing polarization functions. The TZP and TZ2P basis sets result in differences of < 1.25 % in all cases. The largest basis set in a convergence study is usually taken as the reference – QZ4P in this case. From first glance it looks like the triple zeta results are not converged since deviations of up to 3.6 % persist. But a closer look at the QZ4P calculations reveals several issues with linear dependencies leading to a reduced numerical accuracy. Thus, the TZP and TZ2P basis sets with frozen core approximation are considered the most reliable setup.

In a further step, we compared the results from pEDA analysis with the established EDA analysis for the setup derived above (TZ2P(fc) with accuracy 5, see Table 3).

Table 3

The donor-acceptor bonds are discussed first. In both cases, the EDA and pEDA results agree very well with an absolute deviation of < 1% for the energy terms in the upper part of the table. The intermediate terms, $\Delta T^0$ and $\Delta V^0$, exhibit larger differences. This is probably due to the differences in the basis set definitions between the non-periodic and periodic programs used, whereby the latter code neglects some diffuse functions to improve the numerical stability of the program.[52]

The deviations are more significant for the shared-electron bond in ethane. This can be traced back to the different description of the fragments in both approaches. While the EDA approach works with closed-shell fragments, the pEDA uses open-shell fragments directly (see Scheme 3). This leads to a spin-flip error in the EDA which needs to be corrected ($\Delta E_{spin-flip}$) and amounts to -121.3 kJ mol$^{-1}$ for this fragmentation. The improper description of the fragment density has an indirect effect on all analysis terms, leading to higher values for $\Delta E_{Pauli}$, $\Delta E_{elstat}$ and $\Delta E_{orbital}$. Thus, it becomes clear that the usage of spin-unrestricted fragments in the pEDA leads to a more accurate bonding description for shared-electron bonding.[64]

### 4.2.2 Extended Systems

For the extended test systems the results for the convergence w.r.t. basis set size and frozen core approximation are found in Table 4. The convergence w.r.t. k-space sampling and accuracy parameter are summarized in Figure 5 for the examples $H_2$ on Cu (Figure 5a) and $C_2H_2$ on Si (Figure 5b). The underlying numbers and results for the other systems can be found in the Supporting Information.

Table 4

Figure 5

For the basis set study, the findings are comparable to the molecular systems. The frozen core approximation leads to rather small changes in the pEDA terms with few exceptions: The DZ(fc) result for CO on MgO(001) as well as the QZ4P(fc) results for the same system and the $C_2H_2$ on Si(001) case differ significantly from the all electron calculations. For the most relevant TZP and TZ2P basis sets, the saving in computing time (approx. 20%) is well worth the slight errors introduced which are below chemical accuracy (4.2 kJ mol$^{-1}$) for the main pEDA terms.

Regarding the basis set, the TZP and TZ2P results are again very similar for the systems investigated. Surprisingly, even the rather small DZ basis set delivers a very good qualitative picture with the exception of the DZ(fc) results discussed above. While the absolute numbers can differ by up to 15%, most results are much closer to TZP. Care has to be taken when trends are discussed with the DZ basis set, since description of the relative contributions of $\Delta E_{elstat}$ and $\Delta E_{orb}$ can differ in comparison to the larger basis sets. The results for QZ4P are again noteworthy. Initially, it seems that the triple zeta results are not converged yet w.r.t. the larger basis set. A closer look reveals, that the large deviations observed in going from TZ2P to QZ4P (up to 44 kJ mol$^{-1}$ as in the $\Delta E_{orb}$ term for $C_2H_2$ on Si(001)) can be explained again by linear dependencies which show up in the QZ4P calculations and which are indicated in Table 4. As another check for the convergence of the TZ2P results, the $\Delta E_{int}$ values were compared to results from a calculation based on a plane wave basis set shown in the most-right column of Table 4. It becomes clear that for the metallic and the semiconducting system, the TZ2P basis set delivers well converged dissociation energies.

Next, we want to discuss the convergence w.r.t the sampling of reciprocal space as shown in Figure 5. It was recognized before, that the convergence of the pEDA terms for a metallic system can depend strongly on the k-space sampling. This is due to the separate orthogonalization steps for occupied and virtual space and discontinuities can be generated due to change of occupations from one k-point to the next.[35a] This is reproduced in our case, and for the system $H_2$ on Cu (Figure 5a) a tight sampling of more than 15 k-points per Å$^{-1}$ is needed to converge the pEDA terms below an accuracy of 1% relative to the finest k-mesh. Nevertheless, this is not unique to the pED, since it is a general feature of the computation of metallic systems with PBC. Specific for the method here is the different convergence behavior for the pEDA terms. While $\Delta E_{elstat}$ converges very quickly (< 5 k-points per Å$^{-1}$), $\Delta E_{orb}$ and $\Delta E_{Pauli}$ are much more strongly dependent on a good sampling of the Brioullin zone. This has to be taken into account, when bonding trends (relative contributions of these terms) are discussed for a limited k-point sampling. A much quicker convergence is observed for the semiconducting example $C_2H_2$ on Si(001) (Figure 5b). Around 10 k-points per Å$^{-1}$ deliver well converged results and even the sampling at the Γ-point is a very good approximation. A closer look at the underlying numbers (see Table S10) reveals an alternating behavior of the results with the usage of a linear (even k-space settings) and quadratic (uneven) tetrahedron method for the sampling (see Method section for details). As suggested, only results from the same method should be used and the quadratic method delivers more accurate results.[65] For the insulating system CO on MgO(001) it is sufficient to include the Γ-point for converged numbers (except for the high coverage regime, see Table S4).

The accuracy parameter convergence is much less system dependent (Figure 5). For all test systems, the pEDA terms are converged to <1 % for an accuracy setting of 4 or 5. This is confirmed by the numerical results in the Supporting Information with one exception. The results for CO on MgO(001) show a non-monotonous convergence w.r.t the accuracy parameter mostly due to changes in the $\Delta E_{elstat}$ term. It might be concluded that the weak interaction in this system leads to small absolute energy terms which pose a difficulty for deriving well converged analysis results.

The convergence tests for several extended systems thus lead to the conclusion that a triple zeta basis set with frozen core approximation together with an accuracy setting of 4-5 should be used and the k-space sampling has to be adjusted to the electronic structure of the surface investigated. This standard will be employed in the next sub-sections to discuss the chemical bonding in the test systems.

## 4.3  Applications of the pEDA

### 4.3.1  Dissociative adsorption of $H_2$ on Pd(001) and Cu(001)

The dissociative adsorption of $H_2$ is a non-activated process on Pd while an activated process is observed on Cu. This lead to the aforementioned investigation by Philipsen and Baerends (PB) focusing on the role of Pauli repulsion in this adsorption process.[35a] It was concluded that the relief of Pauli repulsion is the decisive factor for the different behavior on Pd and Cu. The difference in our implementation is now that we are able to explicitly calculate the $\Delta E_{elstat}$ and $\Delta E_{Pauli}$ term while the $\Delta E^0$ (= $\Delta E_{elstat}$ + $\Delta E_{Pauli}$) and $\Delta E_{orb}$ term are directly comparable to the analysis by PB. Are our results in agreement with PB and do we gain additional insight by the further decomposition of the steric interaction term?

A series of pEDA analyses was conducted for $H_2$ approaching the surfaces (Figure 1). The resulting energy terms along the reaction coordinate (d(M-$H_2$)) are shown in Figure 6 and the respective numbers can be found in the Supporting Information. The same data in the format used by PB can be found in the SI for direct comparison (Figure S1).

Figure 6

We confirm the finding by PB that the interaction energy is lower, the Pauli repulsion higher and the orbital interaction weaker on Cu compared to Pd along the reaction coordinate (Figure 6a). Additionally, we can now state that the electrostatic stabilization term $\Delta E_{elstat}$ is also more favorable on the Pd surface (Figure 6b). All differences get larger along the reaction coordinate. Thus, we obtain an additional stabilization term on the Pd surface of electrostatic nature which adds to the picture of a non-activated process on this surface for the dissociative adsorption of $H_2$. Nevertheless, the Pauli repulsion stays the leading term as found by PB and thus dominates the pEDA differences between the two surfaces especially at larger distances. The different trends for the surfaces can be understood in terms of electron density at the surface as indicated via the local density of states (LDOS) shown in Figure 7.

Figure 7

The electron density at the Cu surface (Figure 7a) is thereby found to be significantly reduced in comparison to the Pd surface (dark blue areas in Figure 7b). Thus, the approaching $H_2$ molecule interacts with more surface electrons in the latter case which is reflected in higher values for the attractive terms $\Delta E_{orb}$ and $\Delta E_{elstat}$ as well as a higher value for $\Delta E_{Pauli}$ on Pd along the reaction coordinate (Figure 6).

### 4.3.2 Adsorption of CO on MgO(001)

Carbon monoxide is a typical probe molecule for surface science while the MgO(001) surface is a prototype for an insulating surface. Therefore, this system has been widely investigated in the past and among other methods has been analyzed by the CSOV method in a periodic implementation by Hernandez, Zicovich-Wilson and Sanz (HWS).[44b] This analysis confirmed previous findings[66] indicating electrostatic forces and Pauli repulsion as the main driving forces in this weakly bound system. Recently, the adsorption energy in this system was derived by highly accurate ab initio methods and the extrapolated adsorption energy found 21.0 ± 1.0 kJ mol$^{-1}$ is in very good agreement with the most reliable experimental values.[67]

To test our pEDA implementation, we reconsidered the setup by HWS and analyzed CO adsorption on MgO for three different coverages (see Figure 2). The pEDA results are found in Table 5. First of all, the adsorption energy derived is in very good agreement with the above-mentioned high level value of 21.0 kJ mol$^{-1}$ ($\Delta E_{int}$ = -22.2 kJ mol$^{-1}$ for $\theta$ = 0.125). There is certainly a degree of error cancellation for our computational setup (possible error sources being e.g. relaxation, zero-point vibrational energy, thermal corrections, slab setup) but it lends confidence to the pEDA results besides the rather small energy terms. As seen in the previous sub-section, care has to be taken to use well converged computational parameters for such a weakly bound system.

Table 5

The bonding picture given by the pEDA shows domination by a strong Pauli repulsion and electrostatic attraction, while the $\Delta E_{orb}$ term is only approximately a third of the attractive interactions. This agrees well with the CSOV results by HWS which can be quantified by summing up selected pEDA terms to deliver their CSOV counterparts: The sum of $\Delta E_{Pauli}$ and $\Delta E_{elstat}$ can be compared to the frozen orbital (FO) term while the $\Delta E_{orb}$ term corresponds to a sum of the polarisation (Pol) and charge transfer (CT) terms. We see the same trend in the orbital term in both methods – with an absolute difference of approx. 5 kJ mol$^{-1}$ – but a qualitative difference in the FO term. While the CSOV gives a rather constant value of approx. 8 kJ mol$^{-1}$ across the three coverage regimes, the pEDA leads to an decrease of the respective sum in going from $\theta = 0.5$ (17.5 kJ mol$^{-1}$) to $\theta = 0.125$ (4.2 kJ mol$^{-1}$). In consequence, the binding energy (BE) in the CSOV analysis is constant across the three coverages investigated while the pEDA leads to an increase of the interaction energy $\Delta E_{int}$. While a quantitative agreement should not be expected due to the different density functionals used and the neglected contributions outlined above, it is still noteworthy that the trend in our data is consistent with the experimentally observed linear increase of the adsorption energy towards lower coverage.[68] Thus, we conclude that the bonding in CO on MgO(001) is indeed dominated by Pauli repulsion and electrostatics but the relative importance of these terms shifts upon decreasing coverages and results in a higher binding energy in the low coverage regime.

### 4.3.3 Adsorption of CO on Si(001)

Carbon monoxide can also be taken as a probe molecule for semiconductor surfaces. The most investigated and technologically most relevant semiconductor surface is Si(001). This surface exhibits a well-known c(4x2) reconstruction revealing buckled dimers with one Si atom lying above the dimer plane (Si$_{up}$) and one Si atom below the plane (Si$_{down}$).[69] Analysis of the band structure reveals one occupied and one unoccupied surface state.[70] In a localized view on the bonding, this can be understood as the Si$_{up}$ atom exhibiting an electron lone-pair and the Si$_{down}$ atom an empty orbital. Therefore, this surface dimer is prone to electrophilic (Si$_{up}$) or nucleophilic (Si$_{down}$) attack.[71] Thus, the interaction of the CO molecule with the Si$_{down}$ atom can be used as a typical example for a donor acceptor bond on a semiconductor surface (Scheme 4a).

This bonding mode was confirmed by experimental and computational studies.[72] We therefore do not aim at reinvestigating the adsorption behavior of this system. From analysis of vibrational signatures, a covalent bonding character with typical donation/back-donation characteristics of CO was concluded.[72c] This is in agreement with the Dewar-Chatt-Duncanson model for transition metal complexes[73] which is known as the Blyholder model in surface science.[74] How can we quantify this view on the bonding of CO on Si(001)?

Table 6

The structural optimization of one CO molecule on the Si(001)c(4x2) unit cell (resulting in a coverage of $\theta = 0.25$ w.r.t. surface Si atoms) as outlined in the method section leads to a structure which is in quantitative agreement with previous PBE results regarding the bond lengths (Figure 3a).[72c] The results for the bonding analysis on this system are shown in Table 6. The bond dissociation energy ($\Delta E_{bond}$ = -102.3 kJ mol$^{-1}$) is higher compared to the previous finding ($E_{ads} = -\Delta E_{bond}$ = 0.92 eV = 88.8 kJ mol$^{-1}$).[72c] This difference can be understood by the inclusion of dispersion effects in our study ($\Delta E_{disp}$ = 10.7 kJ mol$^{-1}$) and the dissimilarities in the computational setup (e.g. slab size, basis set). The pEDA results now give a rather high Pauli repulsion term ($\Delta E_{Pauli}$ = 855.1 kJ mol$^{-1}$) and near equal contribution of attractive electrostatic (46.0 %) and orbital (54.0 %) terms. This ratio was also found in a recent EDA study on CO binding to a molecular Lewis acid.[75] According to the models discussed above, the main contribution to the orbital term stems from the interaction of the highest occupied molecular orbital (HOMO) of the CO molecule with the unoccupied surface state which is mainly localized at the Si$_{down}$ atom (Figure 3a). The strong overlap between the two orbitals explains the high value for the orbital interaction term.

The rather localized nature of this interaction enables us to analyze this bond additionally in a cluster model. A Si$_{15}$H$_{16}$ cluster cut out from the slab setup bearing one CO molecule is shown in Figure 3b. On the one hand, we can now check the numerical accuracy of the new implementation by comparing pEDA data on the cluster with EDA results. On the other hand we can discuss the influence of including the full environment of the surface in the periodic slab model against the truncated nature of the cluster model.

The EDA and pEDA data for the cluster model are in nearly quantitative agreement (< 1 %) confirming the accuracy of the new method. The comparison of pEDA results for cluster and slab approach shows that the bond of CO to Si(001) can be described rather well with a small cluster model. Nevertheless, a closer look reveals that the dispersion energy term ($\Delta E_{disp}$) and the preparation energy of the surface ($\Delta E_{prep}$(Si)) differ more strongly than the other terms since an extended system is replaced by a zero-dimensional cluster. Another difference can be seen in the orbital/band energies shown in Figure 3. While the HOMO energy for the CO does not depend on the application of PBC, the energy for the lowest unoccupied molecular orbital (LUMO) of the cluster differs significantly from the lowest unoccupied crystal orbital (LUCO) of the surface. This is a relevant difference if one is interested in effects like energy level matching for solar cell optimization.[76]

4.3.4 Adsorption of Acetylene on Si(001)

As a last example, we analyzed the bonding of acetylene to a Si(001)c(4x2) substrate as an example for shared-electron binding between surface and adsorbate (Scheme 4b). This enabled us to test the capabilities of the pEDA to work with unrestricted fragments. This system has been much investigated before experimentally[77] as well as computationally[78] and is a prototype system for surface functionalization of semiconductors. Similar to the previous system, we compared the pEDA results for the periodic system (Figure 4a) to EDA results on a cluster model (Si$_{15}$H$_{16}$, Figure 4b) and pEDA results for the same cluster (Table 7).

Table 7

The adsorption structure shown in Figure 4a reveals two unequal Si-C bond lengths (1.896 and 1.913 Å) which is due to the interactions with the surrounding buckled dimers. The C-C bond of acetylene increases upon adsorption from 1.205 Å to 1.358 Å. These bonding parameters are in good agreement with typical values for a C=C double and two Si-C single bonds in line with the Lewis picture sketched in Scheme 4b.

The bond dissociation energy ($\Delta E_{bond}$) for the three approaches chosen is very similar and in good agreement with literature values.[78] In contrast, the bonding analysis terms show notable differences. As was seen for the molecular case of ethane above, the pEDA exhibits numerically smaller values for all analysis terms compared to the EDA. As in the case of ethane discussed above, this can be traced back to the usage of unrestricted fragments in the new method. The differences are significant with deviations of up to 132.5 kJ mol$^{-1}$ ($\Delta E_{Pauli}$). The largest relative deviations between the two methods are as expected seen in the preparation energy ($\Delta E_{prep}$) of both fragments where the spin-flip error has to be taken into account for the EDA. A recalculation of the cluster approach with the pEDA delivers very similar terms and reveals the intrinsic differences between cluster and slab approach: Higher $\Delta E_{elstat}$ and $\Delta E_{prep}$(Si) terms in the finite model. The relative magnitude of the pEDA terms points toward a higher orbital contribution compared to CO on this substrate (Table 6) and larger terms in all cases due to two bonds being formed at the same time. We see that the preparation energy of the acetylene fragment is very significant, thus the linear adsorbate requires a strong bending and an additional spin-flip to enable binding to the Si surface.

The frontier orbitals for the fragments after the preparation step are shown in Figure 4. For finite and infinite approaches the singly occupied orbitals are set up for two covalent σ-bonds between adsorbate and surface atoms. As seen for the previous system, the orbital energies (and also the energy differences) differ for both approaches which might be relevant for charge-transfer investigations.

## 5   Summary

We present an implementation of the energy decomposition analysis method for periodic systems (pEDA) on the density functional level. This enables the analysis of energy contributions for surface-adsorbate binding in diverse systems. The implementation enables the computation of all energy terms and thus leads to quantification of preparation, dispersion, electrostatic, electron-repulsion and orbital contributions to the chemical bond in periodic systems. We tested the pEDA against the established non-periodic EDA method for molecular systems with prototypical bonding scenarios (donor-acceptor bonding for main group and transition metals, shared-electron bonding). Furthermore, we tested the pEDA for four surface adsorbate complexes representing insulating (CO on MgO(001)), metallic ($H_2$ on M(001), M = Pd, Cu) and semiconducting (CO and $C_2H_2$ on Si(001)c(4x2)) substrates. We confirmed the numerical stability of the pEDA and derived a reliable set of computational parameters (TZ2P(fc) basis set, accuracy 5, k-sampling at the Γ-point for semiconductors). The comparison to previous results showed new valuable insight being gained by the pEDA and the analysis of shared-electron binding reveals the necessity for treating the fragments in an unrestricted Kohn-Sham formalism. The need for a periodic description of the extended systems in contrast to a cluster approach for detailed bonding analysis was shown. The results presented lend confidence, that the pEDA will be a powerful tool for the analysis of chemical bonding in surface and materials science in the future.

**Acknowledgement**

Funding by the Deutsche Forschungsgemeinschaft (DFG) through project TO 715/1-1 and the Research Training Group 1782 is gratefully acknowledged. We thank SCM/Amsterdam for kindly providing a developer's license for the BAND code and support. We thank Prof. Tom Ziegler (Canada) for helpful discussions. Computational resources were provided by HRZ Marburg, LOEWE-CSC Frankfurt and HLR Stuttgart.

We are very grateful to Prof. Gernot Frenking (Marburg) for sparking our interest for the world of chemical bonding unicorns with his endless enthusiasm for the topic.

**Figure Captions**

**Scheme 1.** a) Schematic description of the steps in the energy decomposition analysis (EDA) of a chemical bond between two fragments A and B forming an entity AB and b) the energy terms arising in the respective analysis steps.

**Scheme 2.** Molecular test systems for pEDA: a) $H_3N-BH_3$, b) $Cr(CO)_6$ and c) $C_2H_6$ with fragmentation indicated (only σ-donation shown for b).

**Scheme 3.** Pictorial representation of a) restricted open-shell fragments (EDA) with necessary spin-flip and b) unrestricted open-shell fragments (pEDA) available for c) the analysis of an electron-sharing bond at the example of ethane.

**Scheme 4.** Schematic frontier molecular orbital picture of chemical bonding of a) CO and b) $C_2H_2$ to Si(001)c(4x2).

**Figure 1.** Schematic representation of reaction coordinates chosen for $H_2$ approaching a M(001) surface with M = Cu, Pd.

**Figure 2.** Super cells chosen for CO adsorption on MgO(001) resulting in coverages of a) θ = 0.5, b) θ = 0.25 and c) θ = 0.125 w.r.t the number of Mg adsorption sites.

**Figure 3.** Optimized structures of a) slab and b) cluster representation of CO adsorption on a Si(001)c(4x2) surface with selected bond lengths in Å. The HOMO of the CO molecule is shown as well as the LUMO/LUCO of the silicon surface cluster/slab with orbital energies in eV.

**Figure 4.** Optimized structures of a) slab and b) cluster representation of $C_2H_2$ adsorption on a Si(001)c(4x2) surface with selected bond lengths in Å. The SOMOs of the $C_2H_2$ molecule are shown as well as the SOMOs/SOCOs of the silicon surface cluster/slab with orbital energies in eV.

**Figure 5.** Convergence of pEDA results w.r.t. k-space sampling and accuracy parameter for analysis of a) $H_2$ on Cu(001) and b) $C_2H_2$ on Si(001). The dark grey/light grey area indicates convergence to 1% / 5% deviation from the most accurate setting, respectively.

**Figure 6.** Change of pEDA energy terms a) $\Delta E_{Pauli}$ and $\Delta E_{orb}$ as well as b) $\Delta E_{elstat}$ along the reaction coordinate for $H_2$ adsorption on Cu(001) and Pd(001) surfaces.

**Figure 7.** Intersecting plane of the local density of states (0 - 2 eV below the Fermi level ($E_f$)) perpendicular to the surface cutting through the bridging positions for a) Cu(001) and b) Pd(001).

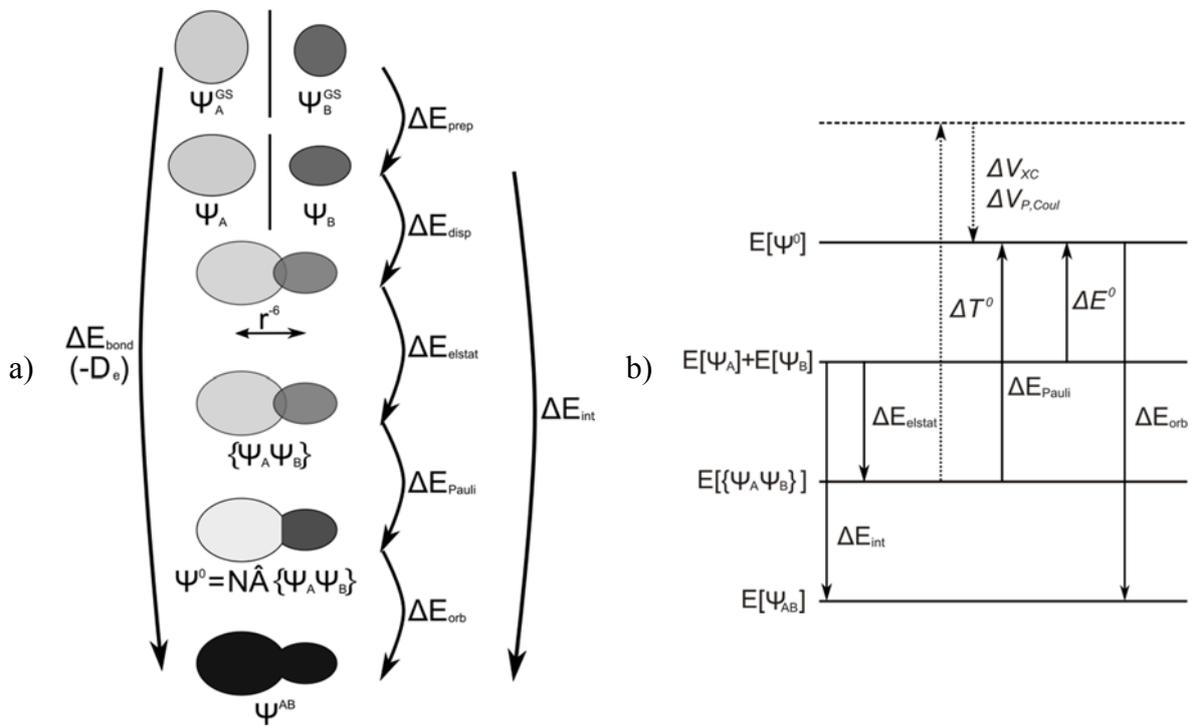

**Scheme 1**

H₃N⟨↑↓⟩ ⟨ ⟩BH₃   OC⟨↑↓⟩ ⟨ ⟩Cr(CO)₅   H₃C⟨↑⟩ ⟨↑⟩CH₃

a)                  b)                   c)

**Scheme 2**

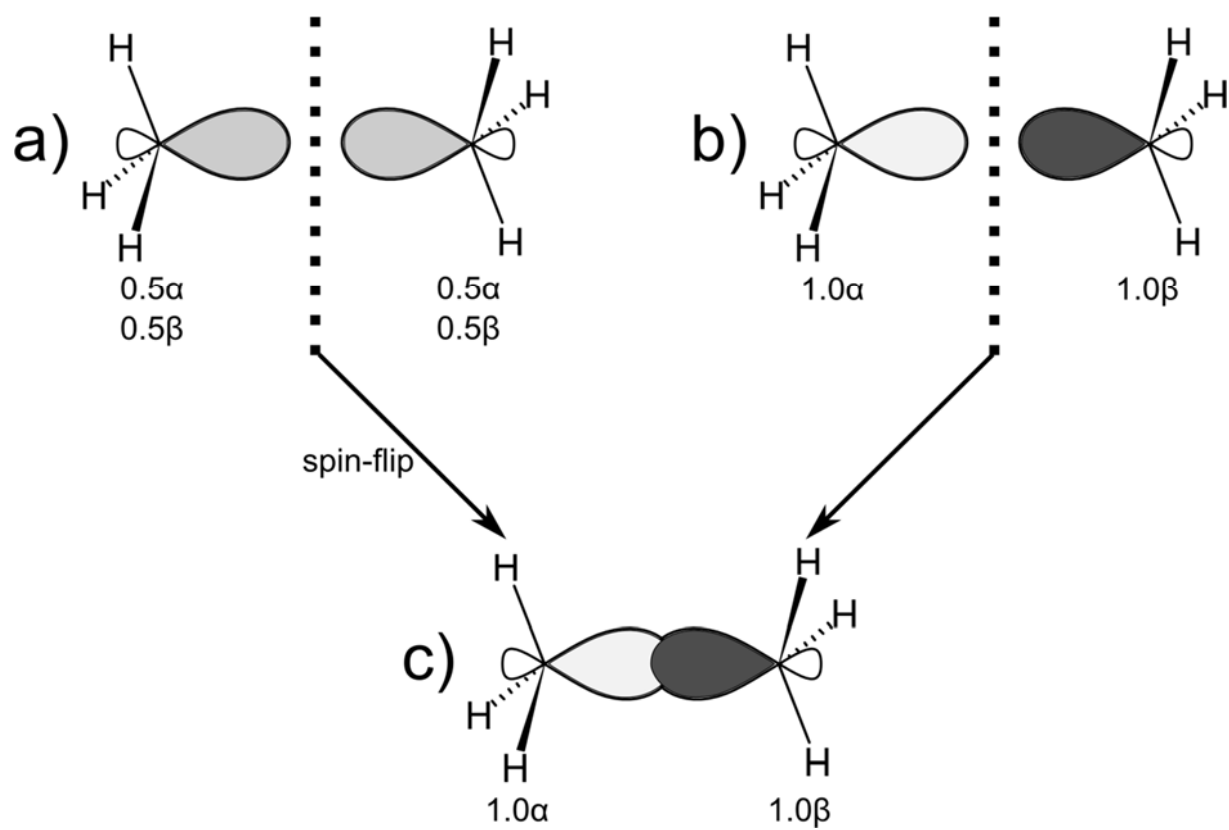

**Scheme 3**

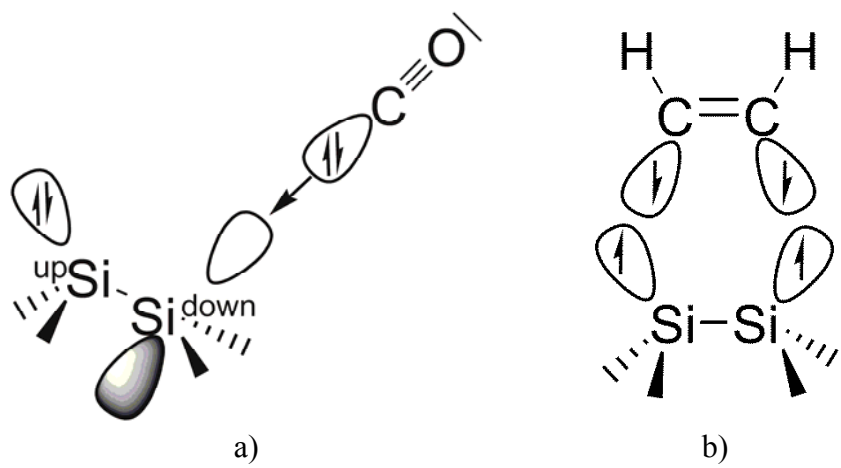

a) b)

**Scheme 4**

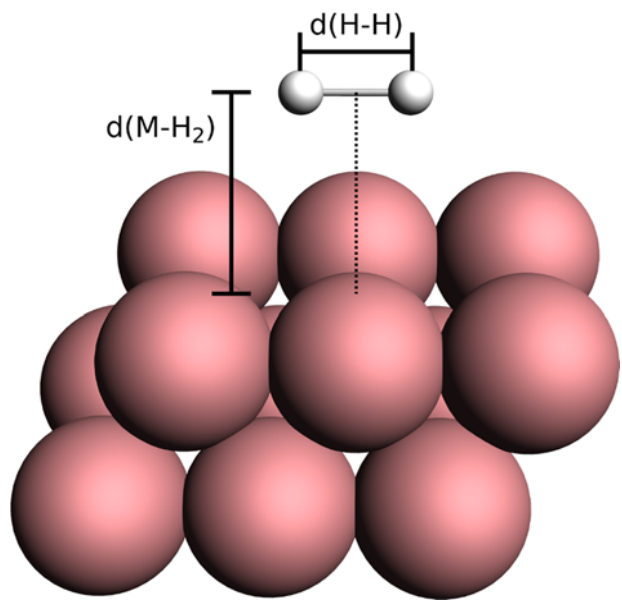

**Figure 1**

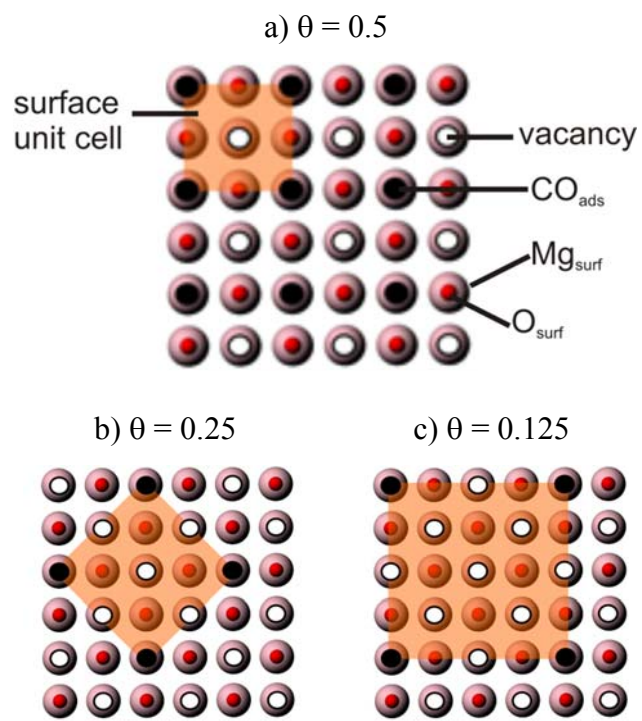

**Figure 2**

a)

b)

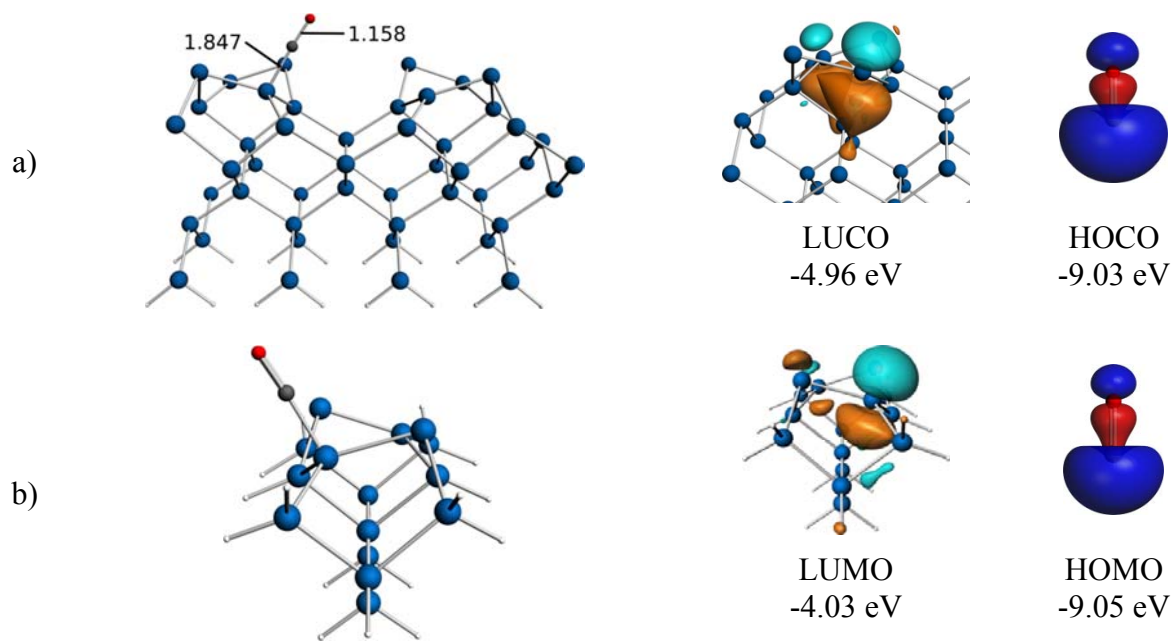

LUCO
-4.96 eV

HOCO
-9.03 eV

LUMO
-4.03 eV

HOMO
-9.05 eV

**Figure 3**

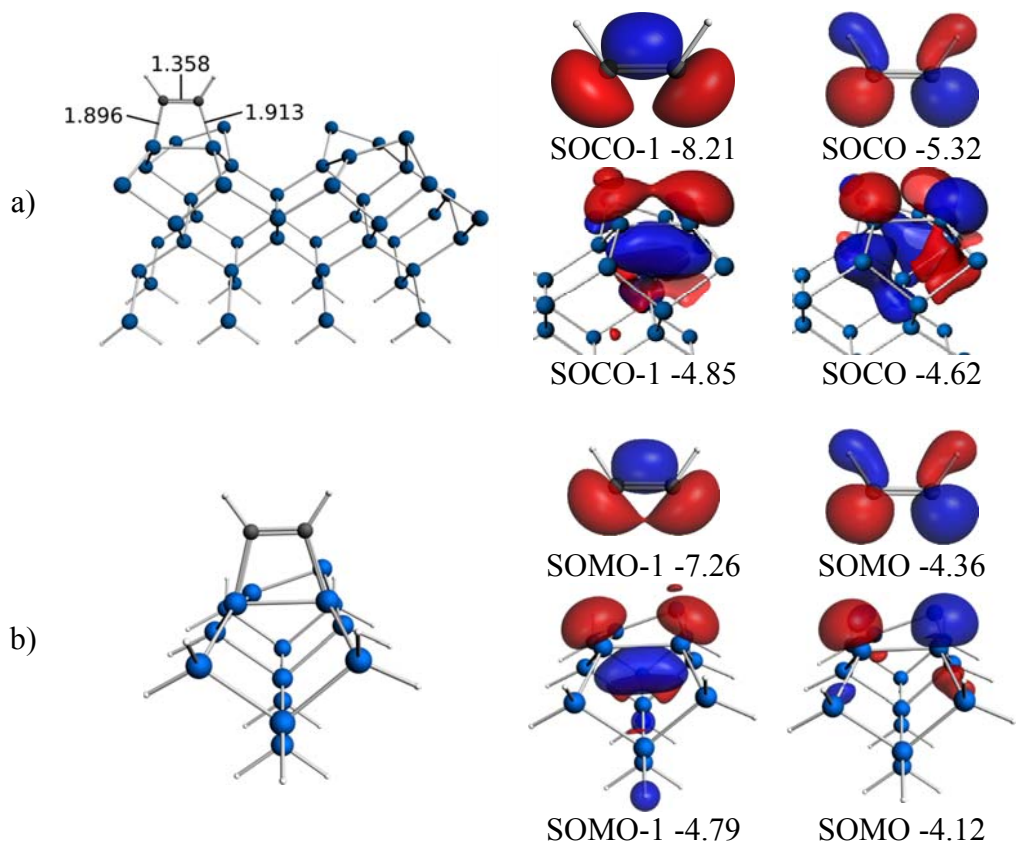

**Figure 4**

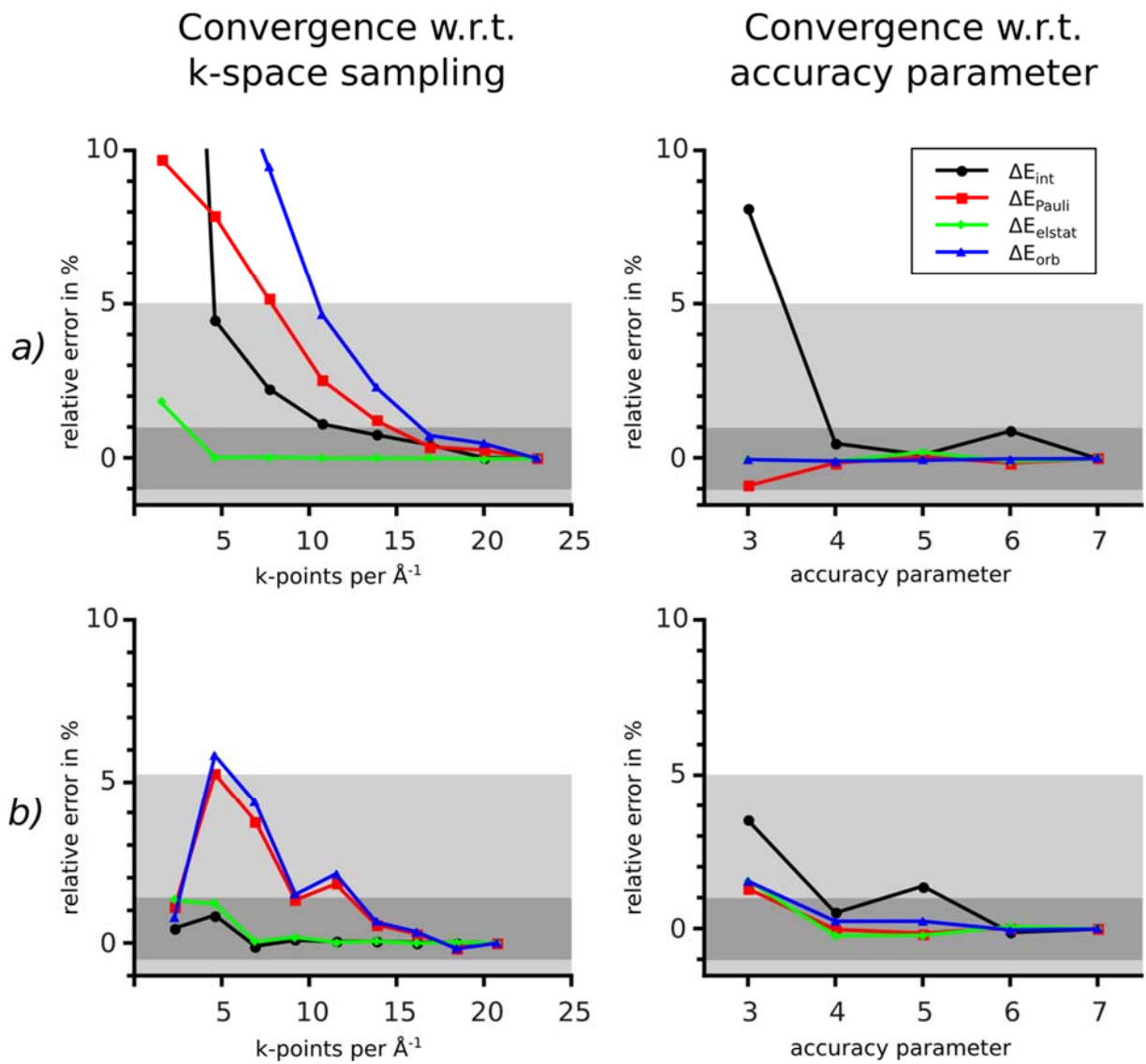

**Figure 5**

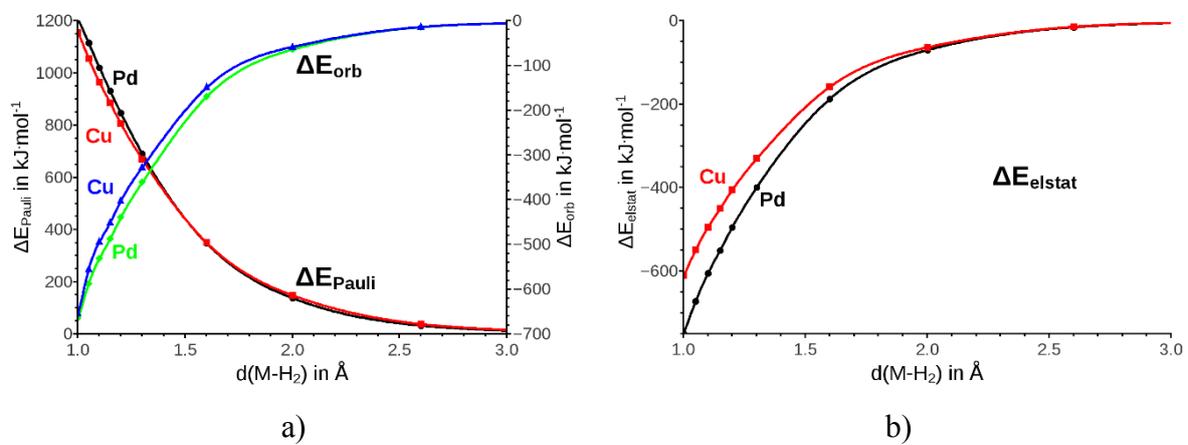

a) b)

**Figure 6**

a)

b)

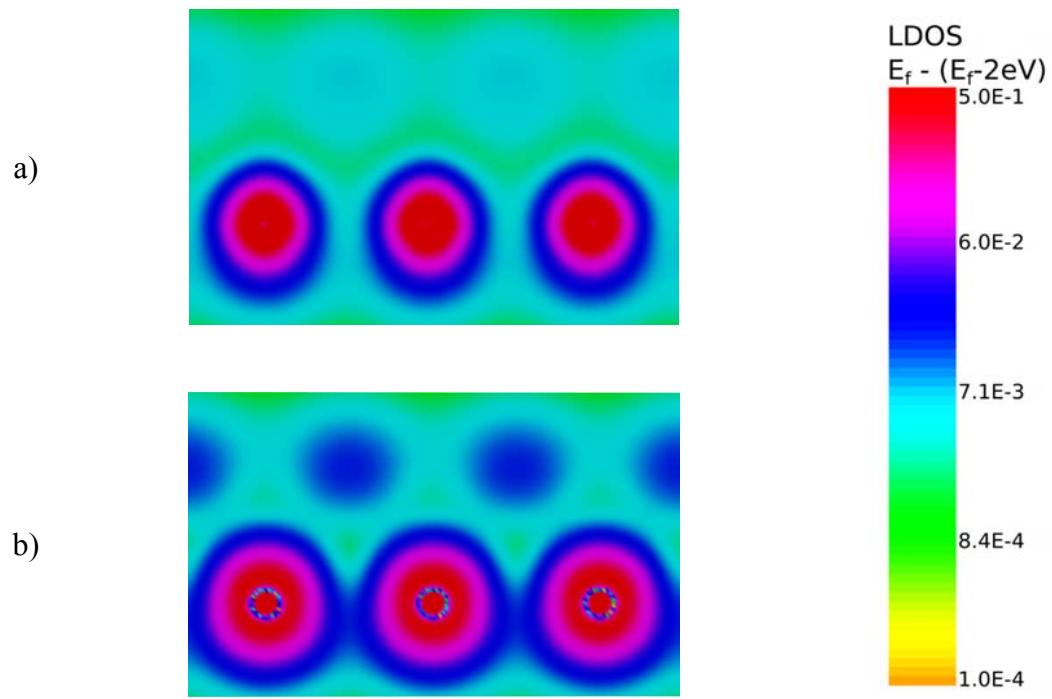

**Figure 7**

**Table 1.** Convergence of pEDA results w.r.t. basis set size and frozen core (fc) approximation for molecular systems.[a]

| basis set | DZ | TZP | TZ2P | QZ4P | | DZ(fc) | TZP(fc) | TZ2P(fc) | QZ4P(fc) |
|---|---|---|---|---|---|---|---|---|---|
| | | | | | $H_3N-BH_3$ | | | | |
| $\Delta E_{int}$ | -221.0 | -187.2 | -187.1 | -188.5 | | -219.6 | -186.6 | -186.6 | -188.1 |
| $\Delta E_{Pauli}$ | 475.6 | 453.0 | 453.5 | 461.3 | | 477.4 | 453.8 | 454.7 | 461.5 |
| $\Delta E_{elstat}$[b] | -444.1 | -326.3 | -323.4 | -322.4 | | -444.3 | -327.9 | -325.1 | -323.0 |
| | (63.7%) | (51.0%) | (50.5%) | (49.6%) | | (63.7%) | (51.2%) | (50.7%) | (49.7%) |
| $\Delta E_{orb}$[b] | -252.5 | -313.9 | -317.2 | -327.4 | | -252.6 | -312.5 | -316.2 | -326.6 |
| | (36.3%) | (49.0%) | (49.5%) | (50.4%) | | (36.3%) | (48.8%) | (49.3%) | (50.3%) |
| $\Delta T^0$ | 1774.9 | 1647.2 | 1646.6 | 1706.8 | | 1794.3 | 1655.6 | 1654.3 | 1706.2 |
| $\Delta V^0$ | -1743.4 | -1520.5 | -1516.5 | -1567.9 | | -1761.2 | -1529.6 | -1524.8 | -1567.8 |
| | | | | | OC-Cr(CO)$_5$ | | | | |
| $\Delta E_{int}$ | -197.6 | -187.8 | -189.1 | -185.4 | | -254.8 | -187.7 | -189.1 | -187.0 |
| $\Delta E_{Pauli}$ | 439.8 | 451.9 | 456.0 | 461.9 | | 442.9 | 452.0 | 456.4 | 458.9 |
| $\Delta E_{elstat}$[b] | -326.1 | -327.4 | -331.1 | -323.0 | | -344.6 | -327.8 | -331.8 | -323.6 |
| | (51.2%) | (51.2%) | (51.3%) | (49.9%) | | (49.4%) | (51.2%) | (51.4%) | (50.1%) |
| $\Delta E_{orb}$[b] | -311.3 | -312.3 | -314.1 | -324.3 | | -353.0 | -311.9 | -313.7 | -322.3 |
| | (48.8%) | (48.8%) | (48.7%) | (50.1%) | | (50.6%) | (48.8%) | (48.6%) | (49.9%) |
| $\Delta T^0$ | 2694.3 | 2558.7 | 2579.1 | 2578.4 | | 2727.4 | 2561.5 | 2583.3 | 2575.2 |
| $\Delta V^0$ | -2580.6 | -2434.1 | -2454.2 | -2439.4 | | -2629.2 | -2437.3 | -2458.7 | -2439.9 |
| | | | | | $H_3C-CH_3$ | | | | |
| $\Delta E_{int}$ | -474.0 | -467.5 | -466.5 | -466.0 | | -472.3 | -467.4 | -466.4 | -465.8 |
| $\Delta E_{Pauli}$ | 793.5 | 751.4 | 752.3 | 769.0 | | 791.9 | 752.1 | 752.7 | 770.3 |
| $\Delta E_{elstat}$[b] | -606.4 | -524.8 | -526.4 | -546.2 | | -606.3 | -525.0 | -526.6 | -547.4 |
| | (47.8%) | (43.1%) | (43.2%) | (44.2%) | | (48.0%) | (43.1%) | (43.2%) | (44.3%) |
| $\Delta E_{orb}$[b] | -661.1 | -694.2 | -692.4 | -688.7 | | -657.9 | -694.5 | -692.6 | -688.8 |
| | (52.2%) | (56.9%) | (56.8%) | (55.8%) | | (52.0%) | (56.9%) | (56.8%) | (55.7%) |
| $\Delta T^0$ | 2142.1 | 1909.6 | 1914.0 | 1951.6 | | 2146.0 | 1907.7 | 1911.1 | 1953.9 |
| $\Delta V^0$ | -1955.0 | -1683.0 | -1688.0 | -1728.8 | | -1960.4 | -1680.6 | -1684.9 | -1730.9 |

[a] All values in kJ mol$^{-1}$ using BP86 with accuracy 5. See Scheme 2 for fragmentation.
[b] The percentage values give the contribution to the total attractive interactions $\Delta E_{elstat} + \Delta E_{orb}$.

**Table 2.** Convergence of pEDA results w.r.t. accuracy parameter for molecular systems.[a]

| accuracy | H$_3$N-BH$_3$ | | | OC-Cr(CO)$_5$ | | | H$_3$C-CH$_3$ | | |
|---|---|---|---|---|---|---|---|---|---|
| | 3 | 5 | 7 | 3 | 5 | 7 | 3 | 5 | 7 |
| $\Delta E_{int}$ | -180.2 | -186.6 | -186.8 | -155.2 | -189.1 | -189.0 | -459.2 | -466.4 | -466.5 |
| $\Delta E_{Pauli}$ | 463.8 | 454.7 | 454.7 | 494.5 | 456.4 | 456.3 | 768.0 | 752.7 | 752.7 |
| $\Delta E_{elstat}$[b] | -327.4 | -325.1 | -325.2 | -335.7 | -331.8 | -331.6 | -536.8 | -526.6 | -526.4 |
| | (50.8%) | (50.7%) | (50.7%) | (51.7%) | (51.4%) | (51.4%) | (43.7%) | (43.2%) | (43.2%) |
| $\Delta E_{orb}$[b] | -316.6 | -316.2 | -316.3 | -314.0 | -313.7 | -313.7 | -690.4 | -692.6 | -692.7 |
| | (49.2%) | (49.3%) | (49.3%) | (48.3%) | (48.6%) | (48.6%) | (56.3%) | (56.8%) | (56.8%) |
| $\Delta T^0$ | 1639.7 | 1654.3 | 1654.9 | 2411.0 | 2583.3 | 2584.0 | 1884.9 | 1911.1 | 1911.5 |
| $\Delta V^0$ | -1503.3 | -1524.8 | -1525.4 | -2252.2 | -2458.7 | -2459.4 | -1653.7 | -1684.9 | -1685.2 |

[a] All values in kJ mol$^{-1}$ using BP86/TZ2P(fc). See Scheme 3 for fragmentation.
[b] The percentage values give the contribution to the total attractive interactions $\Delta E_{elstat} + \Delta E_{orb}$.

**Table 3.** Comparison of EDA and pEDA results for molecular systems.[a]

|  | $H_3N$-$BH_3$ | | OC-Cr(CO)$_5$ | | $H_3C$-$CH_3$ | |
|---|---|---|---|---|---|---|
|  | EDA | pEDA | EDA | pEDA | EDA | pEDA |
| $\Delta E_{int}$ | -191.3 | -191.4 | -193.4 | -195.6 | -484.3 | -470.6 |
| $\Delta E_{disp}$ | -4.8 | -4.8 | -6.5 | -6.5 | -4.2 | -4.2 |
| $\Delta E_{Pauli}$ | 455.4 | 454.7 | 459.1 | 456.4 | 840.5 | 752.7 |
| $\Delta E_{elstat}$[b] | -323.6 | -325.1 | -330.5 | -331.8 | -549.7 | -526.6 |
|  | (50.4%) | (50.7%) | (51.2%) | (51.4%) | (41.6%) | (43.2%) |
| $\Delta E_{orb}$[b] | -318.3 | -316.2 | -315.5 | -313.7 | -770.9 | -692.6 |
|  | (49.6%) | (49.3%) | (48.8%) | (48.6%) | (58.4%) | (56.8%) |
| $\Delta E_{prep}$ | 53.2 | 53.3 | 8.2 | 8.9 | 212.1 | 75.6 |
| $\Delta E_{spin-flip}$[c] |  |  |  |  | -121.3 |  |
| $\Delta E_{bond}$ | -138.1 | -138.1 | -185.2 | -186.7 | -393.5 | -395.0 |
| $\Delta T^0$ | 1681.9 | 1654.3 | 2599.3 | 2579.1 | 2165.5 | 1911.1 |
| $\Delta V^0$ | -1550.1 | -1524.8 | -2470.8 | -2454.2 | -1874.6 | -1684.9 |

[a] All values in kJ mol$^{-1}$ using BP86/TZ2P(fc) with accuracy 5. See Scheme 3 for fragmentation.
[b] The percentage values give the contribution to the total attractive interactions $\Delta E_{elstat} + \Delta E_{orb}$.
[c] Correction term due to the use of restricted fragments in EDA.

**Table 4.** Convergence of pEDA results w.r.t. basis set and frozen core approximation for extended systems investigated. Adsorption energies from plane wave calculations (PW) are given in italics.[a]

| basis set | DZ | TZP | TZ2P | QZ4P | DZ(fc) | TZP(fc) | TZ2P(fc) | QZ4P(fc) | *PW* |
|---|---|---|---|---|---|---|---|---|---|
| | | | | $H_2$@Cu(001)[b] | | | | | |
| $\Delta E_{int}$ | -122.3 | -138.6 | -140.5 | -141.2[c] | -116.2 | -133.8 | -136.4 | -141.6[d] | *-147.1* |
| $\Delta E_{Pauli}$ | 1318.9 | 1303.7 | 1306.5 | 1303.4 | 1318.3 | 1302.0 | 1304.4 | 1299.1 | *(-148.3)*[e] |
| $\Delta E_{elstat}$[c] | -654.5 | -666.7 | -664.9 | -653.4 | -650.8 | -664.1 | -662.7 | -650.7 | |
| | (45.4%) | (46.2%) | (45.9%) | (45.2%) | (45.4%) | (46.3%) | (46.0%) | (45.2%) | |
| $\Delta E_{orb}$[c] | -786.7 | -775.7 | -782.2 | -791.2 | -783.7 | -771.7 | -778.2 | -789.9 | |
| | (54.6%) | (53.8%) | (54.1%) | (54.2%) | (54.6%) | (53.7%) | (54.0%) | (54.8%) | |
| $\Delta T^0$ | 7929.3 | 7994.4 | 8002.3 | 7939.1 | 7911.4 | 7973.2 | 7989.9 | 7923.9 | |
| $\Delta V^0$ | -7265.0 | -7357.3 | -7360.7 | -7289.1 | -7244.0 | -7335.2 | -7348.1 | -7275.5 | |
| | | | | $C_2H_2$@Si(001)[f] | | | | | |
| $\Delta E_{int}$ | -646.4 | -653.3 | -654.6 | -691.5[c] | -651.8 | -653.9 | -655.3 | -663.1[d] | *-527.6* |
| $\Delta E_{Pauli}$ | 1305.4 | 1310.1 | 1312.4 | 1341.4 | 1296.6 | 1306.6 | 1308.3 | 1323.2 | *(-531.8)*[g] |
| $\Delta E_{elstat}$[c] | -876.5 | -820.7 | -820.8 | -842.7 | -877.1 | -821.7 | -821.2 | -843.5 | |
| | (44.9%) | (41.8%) | (41.7%) | (41.5%) | (45.0%) | (41.9%) | (41.8%) | (42.5%) | |
| $\Delta E_{orb}$[c] | -1075.3 | -1142.7 | -1146.2 | -1190.2 | -1071.3 | -1138.8 | -1142.4 | -1142.7 | |
| | (55.1%) | (58.2%) | (58.3%) | (58.5%) | (55.0%) | (58.1%) | (58.2%) | (57.5%) | |
| $\Delta T^0$ | 4165.3 | 3989.4 | 3997.7 | 4061.8 | 4154.9 | 3991.3 | 3998.8 | 4025.7 | |
| $\Delta V^0$ | -3736.4 | -3500.0 | -3506.2 | -3563.1 | -3735.3 | -3506.5 | -3511.7 | -3546.0 | |
| | | | | CO@MgO(001)[h] | | | | | |
| $\Delta E_{int}$ | -28.2 | -22.6 | -22.6 | -10.7[c] | -56.7 | -23.6 | -21.7 | -16.7[d] | *-15.3* |
| $\Delta E_{Pauli}$ | 48.0 | 45.9 | 46.0 | 56.6 | 47.7 | 45.0 | 47.1 | 54.0 | |
| $\Delta E_{elstat}$[c] | -46.7 | -41.3 | -42.2 | -43.0 | -49.6 | -41.5 | -42.0 | -42.3 | |
| | (61.2%) | (60.4%) | (61.4%) | (63.9%) | (47.5%) | (60.6%) | (61.0%) | (59.9%) | |
| $\Delta E_{orb}$[c] | -29.5 | -27.1 | -26.5 | -24.3 | -54.9 | -27.0 | -26.8 | -28.4 | |
| | (38.8%) | (39.6%) | (38.6%) | (36.1%) | (52.5%) | (39.4%) | (39.0%) | (40.1%) | |
| $\Delta T^0$ | 625.8 | 555.2 | 557.2 | 613.2 | 613.6 | 554.6 | 557.7 | 609.9 | |
| $\Delta V^0$ | -624.5 | -550.6 | -553.4 | -599.6 | -615.4 | -551.2 | -552.6 | -598.2 | |

[a] All values in kJ mol$^{-1}$.
[b] BP86, k-space 5, accuracy 5.
[c] The percentage values give the contribution to the total attractive interactions $\Delta E_{elstat} + \Delta E_{orb}$.
[d] Smallest eigenvalue during Löwdin transformation is smaller than $1*10^{-6}$.
[e] $\Delta E_{int}$ using BAND with PBE/TZ2P(fc), k-space 10, accuracy 5 for comparison.
[f] PBE, k-space 1, accuracy 5.
[g] $\Delta E_{int}$ using BAND with PBE/TZP(fc), k-space 7, accuracy 5 and spin restricted fragments for comparison.
[h] PBE, k-space 1, accuracy 5, $\theta = 0.5$.

**Table 5.** Comparison of pEDA and CSOV results for CO on MgO(001).[a]

|  | CSOV[b] | | |  | pEDA | | |
|---|---|---|---|---|---|---|---|
| θ | 0.5 | 0.25 | 0.125 |  | 0.5 | 0.25 | 0.125 |
| BE | -13.5 | -13.9 | -13.5 | $\Delta E_{int}$ | -10.0 | -17.0 | -22.2 |
|  |  |  |  | $\Delta E_{Pauli}$ | 52.1 | 48.8 | 46.6 |
|  |  |  |  | $\Delta E_{elstat}$ | -34.5 | -39.7 | -42.3 |
| FO | 8.1 | 7.8 | 7.8 | $\Delta E_{Pauli} + \Delta E_{elstat}$ | 17.6 | 9.1 | 4.3 |
| Pol | -3.4 | -3.3 | -4.1 |  |  |  |  |
| CT | -18.2 | -18.4 | -17.2 |  |  |  |  |
| Pol + CT | -21.6 | -21.7 | -21.3 | $\Delta E_{orb}$ | -27.6 | -26.1 | -26.4 |

[a] All values in kJ mol$^{-1}$. See Figure 2 for definition of coverage. PBE, TZ2P(fc), accuracy 5, k-space 4.
[b] Values taken from Ref. 43b. See text for definition of terms.

**Table 6.** Comparison of cluster and slab approach with EDA and pEDA for CO on Si(001).[a]

|  | Cluster | | Slab |
|---|---|---|---|
|  | EDA[b] | pEDA[b] | pEDA[c] |
| $\Delta E_{int}$    | -106.8   | -105.7   | -113.5   |
| $\Delta E_{disp}$   | -7.6     | -7.6     | -10.7    |
| $\Delta E_{Pauli}$  | 838.7    | 836.7    | 855.1    |
| $\Delta E_{elstat}$[d] | -432.7 | -432.7 | -440.7 |
|                     | (46.1%)  | (46.3%)  | (46.0%)  |
| $\Delta E_{orb}$[d] | -505.2   | -502.2   | -517.3   |
|                     | (53.9%)  | (53.7%)  | (54.0%)  |
| $\Delta E_{prep}$(CO) | 2.4    | 2.4      | 2.4      |
| $\Delta E_{prep}$(Si) | 3.8    | 4.7      | 11.3     |
| $\Delta E_{bond}$   | -100.5   | -98.7    | -99.9    |
| $\Delta T^0$        | 2835.0   | 2830.1   | 2911.0   |
| $\Delta V^0$        | -2429.0  | -2426.1  | -2496.6  |

[a] All values in kJ mol$^{-1}$. See Scheme 4a for fragmentation.
[b] PBE/TZ2P, accuracy 6.
[c] PBE/TZ2P, accuracy 6, k-space 4.
[d] The percentage values give the contribution to the total attractive interactions $\Delta E_{elstat} + \Delta E_{orb}$.

**Table 7.** Comparison of cluster and slab approach with EDA and pEDA for $C_2H_2$ on Si(001).[a]

|  | Cluster | | Slab |
|---|---|---|---|
|  | EDA[b] | pEDA[b] | pEDA[c] |
| $\Delta E_{int}$ | -707.5 | -688.6 | -667.5 |
| $\Delta E_{disp}$ | -11.3 | -11.3 | -12.2 |
| $\Delta E_{Pauli}$ | 1440.8 | 1302.4 | 1308.3 |
| $\Delta E_{elstat}$[d] | -905.3 | -855.2 | -821.2 |
|  | (42.4%) | (43.2%) | (41.8%) |
| $\Delta E_{orb}$[d] | -1231.7 | -1124.5 | -1142.4 |
|  | (57.6%) | (56.8%) | (58.2%) |
| $\Delta E_{prep}(C_2H_2)$ | 501.3 | 364.2 | 364.4 |
| $\Delta E_{spin-flip}(C_2H_2)$[e] | -123.6 |  |  |
| $\Delta E_{prep}(Si)$ | 60.5 | 30.4 | 5.7 |
| $\Delta E_{spin-flip}(Si)$[e] | -31.3 |  |  |
| $\Delta E_{bond}$ | -300.6 | -294.0 | -297.4 |
| $\Delta T^0$ | 4344.0 | 3911.2 | 3998.8 |
| $\Delta V^0$ | -3808.5 | -3464.0 | -3511.7 |

[a] All values in kJ mol$^{-1}$. See Scheme 4b for fragmentation.
[b] PBE/TZ2P(fc), accuracy 5.
[c] PBE/TZ2P(fc), accuracy 5, k-space 1.
[d] The percentage values give the contribution to the total attractive interactions $\Delta E_{elstat} + \Delta E_{orb}$.
[e] Correction term due to the use of restricted fragments in EDA.

Supporting Information for

# A periodic Energy Decomposition Analysis (pEDA) method for the Investigation of Chemical Bonding in Extended Systems

Marc Raupach and Ralf Tonner

**Table S1**. Convergence of pEDA results w.r.t. **accuracy** parameter for $H_2$ on Cu(001).[a]

| accuracy | 3 | 4 | 5 | 6 | 7 |
|---|---|---|---|---|---|
| $\Delta E_{int}$ | -147.3 | -136.9 | -136.4 | -137.5 | -136.3 |
| $\Delta E_{Pauli}$ | 1292.2 | 1301.8 | 1304.4 | 1301.7 | 1303.6 |
| $\Delta E_{elstat}$ | -661.1 | -660.8 | -662.7 | -660.7 | -661.3 |
| $\Delta E_{orb}$ | -778.3 | -777.9 | -778.2 | -778.5 | -778.6 |
| $\Delta T^0$ | 7993.7 | 7956.9 | 7989.9 | 7993.3 | 7994.5 |
| $\Delta V^0$ | -7362.7 | -7316.0 | -7348.1 | -7352.2 | -7352.2 |

[a] All values in kJ mol$^{-1}$. BP86, k-space 5, TZ2P(fc).

**Table S2**. Convergence of pEDA results w.r.t. **k-space sampling** for $H_2$ on Cu(001).[a]

| k-space | 1 | 3 | 5 | 7 | 9 | 11 | 13 | 15 |
|---|---|---|---|---|---|---|---|---|
| $\Delta E_{int}$ | -194.0 | -136.7 | -133.8 | -132.3 | -131.8 | -131.4 | -130.8 | -130.8 |
| $\Delta E_{Pauli}$ | 1358.2 | 1335.5 | 1302.0 | 1269.4 | 1253.3 | 1242.7 | 1241.4 | 1238.0 |
| $\Delta E_{elstat}$ | -675.8 | -664.0 | -664.1 | -663.9 | -663.8 | -663.9 | -663.8 | -663.8 |
| $\Delta E_{orb}$ | -876.3 | -808.1 | -771.7 | -737.8 | -721.3 | -710.3 | -708.5 | -705.0 |
| $\Delta T^0$ | 7332.7 | 7979.3 | 7973.1 | 8115.4 | 8179.8 | 8223.9 | 8245.9 | 8250.8 |
| $\Delta V^0$ | -6650.4 | -7307.9 | -7335.2 | -7509.9 | -7590.4 | -7645.0 | -7668.3 | -7676.6 |

[a] All values in kJ mol$^{-1}$. BP86, accuracy 5, TZP(fc).

**Table S3**. pEDA results for $H_2$ on Cu(001) and Pd(001) for different metal-$H_2$ distances.[a]

| d(Cu-$H_2$) | 0.80 | 0.85 | 0.90 | 0.95 | 1.00 | 1.05 | 1.10 | 1.15 | 1.20 | 1.30 | 1.60 |
|---|---|---|---|---|---|---|---|---|---|---|---|
| $\Delta E_{int}$ | -251.0 | -224.1 | -167.2 | -138.4 | -109.8 | -47.4 | -22.8 | -14.9 | -0.9 | 12.5 | 43.4 |
| $\Delta E_{Pauli}$ | 1518.8 | 1429.5 | 1340.6 | 1248.9 | 1156.0 | 1056.5 | 965.3 | 885.7 | 806.5 | 669.7 | 350.2 |
| $\Delta E_{elstat}$ | -822.0 | -772.2 | -719.9 | -666.6 | -611.4 | -548.9 | -494.5 | -449.9 | -405.1 | -330.4 | -158.0 |
| $\Delta E_{orb}$ | -947.8 | -881.4 | -787.9 | -720.8 | -654.4 | -555.0 | -493.6 | -450.7 | -402.2 | -326.8 | -148.7 |

| d(Pd-$H_2$) | 0.80 | 0.85 | 0.90 | 0.95 | 1.00 | 1.05 | 1.10 | 1.15 | 1.20 | 1.30 | 1.60 |
|---|---|---|---|---|---|---|---|---|---|---|---|
| $\Delta E_{int}$ | -347.1 | -320.8 | -266.4 | -236.4 | -206.1 | -143.5 | -115.4 | -105.9 | -86.7 | -68.5 | -7.9 |
| $\Delta E_{Pauli}$ | 1531.7 | 1451.4 | 1383.8 | 1297.5 | 1207.8 | 1115.4 | 1020.9 | 931.1 | 847.1 | 690.9 | 347.9 |
| $\Delta E_{elstat}$ | -998.1 | -938.5 | -878.0 | -814.9 | -749.5 | -671.9 | -605.6 | -550.1 | -494.6 | -399.0 | -186.7 |
| $\Delta E_{orb}$ | -880.7 | -833.7 | -772.2 | -719.1 | -664.4 | -587.1 | -530.7 | -486.9 | -439.2 | -360.5 | -169.2 |

[a] All values in kJ mol$^{-1}$. BP86, accuracy 5, k-space 5, TZP(fc).

**Table S4**. Convergence of pEDA results w.r.t. **k-space sampling** for CO on MgO(001) for different coverages.[a]

| | $\theta = 0.5$ | | | | |
|---|---|---|---|---|---|
| k-space | 1 | 2 | 3 | 4 | 5 |
| $\Delta E_{int}$ | 15.7 | -11.2 | -9.7 | -10.0 | -10.0 |
| $\Delta E_{Pauli}$ | 77.3 | 51.7 | 52.3 | 52.1 | 52.0 |
| $\Delta E_{elstat}$ | -41.3 | -34.6 | -34.5 | -34.5 | -34.5 |
| $\Delta E_{orb}$ | -20.3 | -28.4 | -27.4 | -27.6 | -27.6 |
| $\Delta T^0$ | 801.1 | 558.7 | 574.9 | 571.2 | 571.0 |
| $\Delta V^0$ | -765.2 | -541.5 | -557.1 | -553.6 | -553.4 |

| | $\theta = 0.25$ | | | | |
|---|---|---|---|---|---|
| k-space | 1 | 2 | 3 | 4 | 5 |
| $\Delta E_{int}$ | -16.2 | -17.3 | -16.9 | -17.0 | -16.9 |
| $\Delta E_{Pauli}$ | 50.8 | 48.4 | 49.0 | 48.8 | 48.8 |
| $\Delta E_{elstat}$ | -40.3 | -39.6 | -39.8 | -39.7 | -39.4 |
| $\Delta E_{orb}$ | -26.7 | -26.1 | -26.1 | -26.1 | -26.3 |
| $\Delta T^0$ | 575.3 | 559.4 | 564.8 | 563.6 | 563.3 |
| $\Delta V^0$ | -564.8 | -550.6 | -555.6 | -554.5 | -554.0 |

| | $\theta = 0.125$ | | | | |
|---|---|---|---|---|---|
| k-space | 1 | 2 | 3 | 4 | 5 |
| $\Delta E_{int}$ | -22.6 | -22.3 | -22.2 | -22.2 | -22.2 |
| $\Delta E_{Pauli}$ | 46.0 | 46.4 | 46.6 | 46.6 | 46.6 |
| $\Delta E_{elstat}$ | -42.2 | -42.3 | -42.3 | -42.3 | -42.3 |
| $\Delta E_{orb}$ | -26.5 | -26.5 | -26.4 | -26.4 | -26.4 |
| $\Delta T^0$ | 557.2 | 560.5 | 561.9 | 561.7 | 561.7 |
| $\Delta V^0$ | -553.4 | -556.4 | -557.6 | -557.5 | -557.5 |

[a] All values in kJ mol$^{-1}$. PBE, TZ2P(fc), accuracy 5.

**Table S5.** Convergence of pEDA results w.r.t. **accuracy** parameter for CO on MgO(001).[a]

| accuracy | 3 | 4 | 5 | 6 | 7 | 8 | 9 |
|---|---|---|---|---|---|---|---|
| $\Delta E_{int}$ | 11.1 | -26.7 | -22.6 | -14.7 | -20.2 | -19.2 | -16.0 |
| $\Delta E_{Pauli}$ | 86.3 | 42.3 | 46.0 | 53.8 | 48.3 | 49.2 | 52.5 |
| $\Delta E_{elstat}$ | -45.5 | -42.0 | -42.2 | -42.3 | -42.3 | -42.2 | -42.3 |
| $\Delta E_{orb}$ | -29.7 | -26.9 | -26.5 | -26.1 | -26.2 | -26.2 | -26.2 |
| $\Delta T^0$ | 578.3 | 566.6 | 557.2 | 564.4 | 560.3 | 560.8 | 564.1 |
| $\Delta V^0$ | -537.5 | -566.3 | -553.4 | -553.0 | -554.3 | -553.8 | -553.8 |

[a] All values in kJ mol$^{-1}$. PBE, k-space 1, TZ2P(fc), ($2\sqrt{2} \times 2\sqrt{2}$) super cell.

**Table S6.** Convergence of pEDA results w.r.t. **k-space sampling** for CO on Si(001).[a]

| k space | 1 | 2 | 3 | 4 | 5 |
|---|---|---|---|---|---|
| $\Delta E_{int}$ | -115.8 | -111.7 | -103.4 | -105.3 | -105.2 |
| $\Delta E_{Pauli}$ | 856.2 | 841.8 | 859.1 | 855.5 | 855.7 |
| $\Delta E_{elstat}$[a] | -443.7 | -434.8 | -442.5 | -440.9 | -441.0 |
| $\Delta E_{orb}$[a] | -528.3 | -518.7 | -520.0 | -519.9 | -519.9 |
| $\Delta T^0$ | 2933.4 | 2887.5 | 2933.8 | 2924.2 | 2924.7 |
| $\Delta V^0$ | -2520.9 | -2480.4 | -2517.2 | -2509.6 | -2510.0 |

[a] All values in kJ mol$^{-1}$. PBE, TZ2P(fc), accuracy 5.

**Table S7.** Convergence of pEDA results w.r.t. **basis set and frozen core** setting for CO on Si(001).[a]

| basis set | DZ | TZP | TZ2P | QZ4P | DZ(fc) | TZP(fc) | TZ2P(fc) | QZ4P(fc) | *PAW* |
|---|---|---|---|---|---|---|---|---|---|
| $\Delta E_{int}$ | -97.8 | -112.8 | -115.4 | -147.1 | -140.2 | -113.4 | -115.8 | -120.5 | *-104.0*[b] |
| $\Delta E_{Pauli}$ | 846.3 | 852.1 | 856.3 | 898.1 | 847.7 | 851.4 | 856.2 | 873.8 | (-103.6)[c] |
| $\Delta E_{elstat}$ | -452.9 | -440.2 | -442.4 | -432.4 | -467.7 | -441.4 | -443.7 | -435.4 | |
| $\Delta E_{orb}$ | -491.2 | -524.7 | -529.3 | -612.7 | -520.2 | -523.4 | -528.3 | 558.9 | |
| $\Delta T^0$ | 2975.0 | 2901.6 | 2923.0 | 3093.7 | 2960.4 | 2909.7 | 2933.4 | 3061.2 | |
| $\Delta V^0$ | -2581.6 | -2489.6 | -2509.1 | -2628.1 | -2580.4 | -2499.6 | -2521.0 | -2622.8 | |

[a] All values in kJ mol$^{-1}$. PBE, k-space 1, accuracy 5.
[b] Interaction energy derived with VASP according to procedure outlined in the method section.
[c] k-space 4, TZ2P(fc) with BAND.

**Table S8.** Convergence of pEDA results w.r.t. **accuracy** parameter for CO on Si(001).[a]

| accuracy | 3 | 4 | 5 | 6 | 7 |
|---|---|---|---|---|---|
| $\Delta E_{int}$ | -116.2 | -115.8 | -115.8 | -114.4 | -114.4 |
| $\Delta E_{Pauli}$ | 868.1 | 856.0 | 856.2 | 857.2 | 857.1 |
| $\Delta E_{elstat}$ | -450.0 | -443.6 | -443.7 | -445.0 | -444.7 |
| $\Delta E_{orb}$ | -534.3 | -528.2 | -528.3 | -526.7 | -526.9 |
| $\Delta T^0$ | 2901.4 | 2925.2 | 2933.4 | 2916.9 | 2915.8 |
| $\Delta V^0$ | -2483.3 | -2512.8 | -2521.0 | -2504.6 | -2503.4 |

[a] All values in kJ mol$^{-1}$. PBE, k-space 1, TZ2P(fc).

**Table S9.** Convergence of pEDA results w.r.t. **accuracy** parameter for $C_2H_2$ on Si(001).[a]

| accuracy | 3 | 4 | 5 | 6 | 7 |
|---|---|---|---|---|---|
| $\Delta E_{int}$ | -656.6 | -653.4 | -655.3 | -655.1 | -653.4 |
| $\Delta E_{Pauli}$ | 1308.4 | 1310.4 | 1308.3 | 1305.2 | 1306.3 |
| $\Delta E_{elstat}$ | -824.5 | -823.4 | -821.2 | -820.7 | -821.3 |
| $\Delta E_{orb}$ | -1140.5 | -1140.4 | -1142.4 | -1139.6 | -1138.3 |
| $\Delta T^0$ | 3909.2 | 3978.4 | 3998.8 | 3974.4 | 3969.0 |
| $\Delta V^0$ | -3425.3 | -3491.4 | -3511.7 | -3489.8 | -3484.0 |

[a] All values in kJ mol$^{-1}$. PBE, k-space 1, TZ2P(fc).

**Table S10.** Convergence of pEDA results w.r.t. **k-space sampling** for $C_2H_2$ on Si(001).[a]

| k-space | 1 | 2 | 3 | 4 | 5 | 6 | 7 | 8 | 9 |
|---|---|---|---|---|---|---|---|---|---|
| $\Delta E_{int}$ | -534.2 | -536.3 | -531.2 | -532.2 | -532.0 | -532.0 | -531.8 | -531.9 | -531.7 |
| $\Delta E_{Pauli}$ | 2104.3 | 2190.3 | 2159.7 | 2109.0 | 2119.3 | 2093.2 | 2087.4 | 2078.4 | 2081.3 |
| $\Delta E_{elstat}$ [a] | -844.1 | -843.2 | -833.5 | -834.7 | -833.4 | -833.6 | -833.2 | -833.4 | -833.2 |
| $\Delta E_{orb}$ [a] | -1794.4 | -1883.4 | -1857.4 | -1806.6 | -1817.9 | -1791.6 | -1786.2 | -1776.8 | -1779.9 |
| $\Delta T^0$ | 5969.9 | 6219.5 | 6157.0 | 6019.2 | 6051.5 | 5979.7 | 5965.0 | 5937.6 | 5947.7 |
| $\Delta V^0$ | -4709.7 | -4872.5 | -4830.8 | -4744.8 | -4765.6 | -4720.2 | -4710.7 | -4692.7 | -4699.5 |

[a] All values in kJ mol$^{-1}$. PBE, , TZP(fc), accuracy 3.

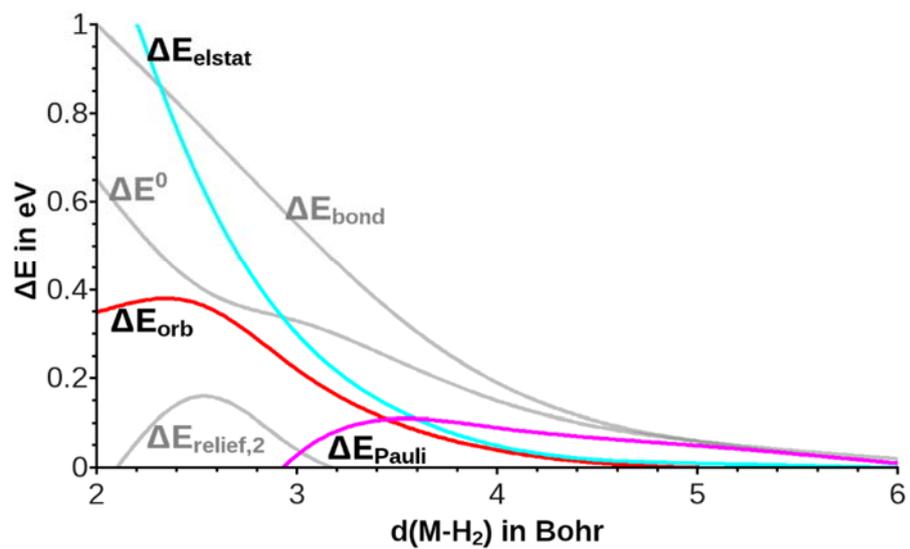

**Figure S1**. Change of EDA energy terms along the reaction coordinate for $H_2$ adsorption on Cu(001) and Pd(001) surfaces in the format used in Ref. 44a for direct comparison.

**Table S11.** Cartesian coordinates (in Å) and formation energies (in kJ mol$^{-1}$) of molecular systems.

| | | $H_3N-BH_3$ | | |
|---|---|---|---|---|
| Formation energy | Atom | x | y | z |
| -3518.22 | N | 0.1954 | 0.0580 | 0.1636 |
| | H | 0.4395 | 0.1997 | 1.1465 |
| | H | -0.8240 | 0.0102 | 0.1002 |
| | H | 0.5676 | -0.8508 | -0.1210 |
| | B | 0.7979 | 1.2753 | -0.7844 |
| | H | 0.4455 | 0.9961 | -1.9139 |
| | H | 0.2877 | 2.2906 | -0.3521 |
| | H | 2.0022 | 1.2296 | -0.6248 |

| | | $Cr(CO)_6$ | | |
|---|---|---|---|---|
| Formation energy | Atom | x | y | z |
| -10026.86 | Cr | 0.0000 | 0.0000 | 0.0000 |
| | C | 1.8933 | 0.1669 | -0.0793 |
| | C | 0.0590 | 0.2275 | 1.8878 |
| | C | 0.1750 | -1.8813 | 0.2213 |
| | C | -0.0589 | -0.2276 | -1.8877 |
| | C | -1.8933 | -0.1668 | 0.0794 |
| | C | -0.1751 | 1.8812 | -0.2214 |
| | O | 3.0397 | 0.2682 | -0.1273 |
| | O | -0.2811 | 3.0204 | -0.3555 |
| | O | -3.0398 | -0.2677 | 0.1275 |
| | O | 0.0948 | 0.3651 | 3.0308 |
| | O | 0.2810 | -3.0205 | 0.3553 |
| | O | -0.0946 | -0.3654 | -3.0308 |

| | | Ethane | | |
|---|---|---|---|---|
| Formation energy | Atom | x | y | z |
| -3860.26 | C | 0.0000 | 0.0000 | -1.0099 |
| | H | -0.2228 | 0.9986 | -1.4109 |
| | H | 0.9762 | -0.3063 | -1.4109 |
| | H | -0.7534 | -0.6923 | -1.4109 |
| | C | 0.0000 | 0.0000 | 0.5216 |
| | H | 0.2226 | -0.9987 | 0.9225 |
| | H | 0.7535 | 0.6921 | 0.9225 |
| | H | -0.9762 | 0.3065 | 0.9225 |

**Table S12**. Cartesian coordinates (in Å) and formation energies (in kJ mol$^{-1}$) for $H_2$ on Cu(001) and Pd(001).

| | | | | Cu+$H_2$ | | |
|---|---|---|---|---|---|---|
| d(Cu-H) | d(H-H) | Formation energy [a] | Atom | x | y | z |
| 1.60 | 0.78 | -3890.99 | Cu | 0.0000 | 0.0000 | 0.0000 |
| 1.50 | 0.82 | -3887.87 | Cu | 1.2763 | -1.2763 | -1.8050 |
| 1.30 | 0.96 | -3873.34 | Cu | 0.0000 | -2.5526 | 0.0000 |
| 1.20 | 1.00 | -3870.21 | Cu | 1.2763 | 1.2763 | -1.8050 |
| 1.10 | 1.06 | -3866.51 | Cu | 2.5526 | -2.5526 | 0.0000 |
| 1.05 | 1.12 | -3864.13 | Cu | -3.8289 | 1.2763 | -1.8050 |
| 1.00 | 1.26 | -3864.10 | Cu | 2.5526 | 0.0000 | 0.0000 |
| 0.95 | 1.32 | -3865.13 | Cu | -3.8289 | -1.2763 | -1.8050 |
| 0.90 | 1.38 | -3867.28 | Cu | -2.5526 | -2.5526 | 0.0000 |
| 0.85 | 1.50 | -3875.43 | Cu | -1.2763 | 1.2763 | -1.8050 |
| 0.80 | 1.66 | -3880.04 | Cu | -2.5526 | 0.0000 | 0.0000 |
| | | | Cu | -1.2763 | -1.2763 | -1.8050 |
| | | | H | (d(H-H))/2 | -1.2763 | d(Cu-$H_2$) |
| | | | H | -(d(H-H))/2 | -1.2763 | d(Cu-$H_2$) |
| | | | Vector 1 | 7.6579 | 0.0000 | |
| | | | Vector 2 | 0.0000 | 5.1053 | |

| | | | | Pd+$H_2$ | | |
|---|---|---|---|---|---|---|
| d(Pd-H) | d(H-H) | Formation energy [a] | Atom | x | y | z |
| 1.60 | 0.78 | -3031.98 | Pd | 0.0000 | 0.0000 | 0.0000 |
| 1.50 | 0.82 | -3036.25 | Pd | 1.3983 | -1.3988 | -1.9775 |
| 1.30 | 0.96 | -3037.74 | Pd | 0.0000 | -2.7966 | 0.0000 |
| 1.20 | 1.00 | -3039.40 | Pd | 1.3983 | 1.3983 | -1.9775 |
| 1.10 | 1.06 | -3039.73 | Pd | 2.7966 | -2.7966 | 0.0000 |
| 1.05 | 1.12 | -3038.75 | Pd | -4.1949 | 1.3983 | -1.9775 |
| 1.00 | 1.26 | -3038.23 | Pd | 2.7966 | 0.0000 | 0.0000 |
| 0.95 | 1.32 | -3040.68 | Pd | -4.1949 | -1.3983 | -1.9775 |
| 0.90 | 1.38 | -3044.94 | Pd | -2.7966 | -2.7966 | 0.0000 |
| 0.85 | 1.50 | -3049.80 | Pd | -1.3983 | 1.3983 | -1.9775 |
| 0.80 | 1.66 | -3053.45 | Pd | -2.7966 | 0.0000 | 0.0000 |
| | | | Pd | -1.3983 | -1.3983 | -1.9775 |
| | | | H | (d(H-H))/2 | -1.3983 | d(Pd-$H_2$) |
| | | | H | -(d(H-H))/2 | -1.3983 | d(Pd-$H_2$) |
| | | | Vector 1 | 8.3900 | 0.0000 | |
| | | | Vector 2 | 0.0000 | 5.5933 | |

[a] Energy in kJ mol$^{-1}$. BP86, TZP(fc), accuracy 5, k-space 10.

**Table S13.** Cartesian coordinates (in Å) and formation energies (in kJ mol$^{-1}$) for CO on MgO for three different coverages θ.

| | | θ = 0.5 | | |
|---|---|---|---|---|
| Formation energy | Atom | x | y | z |
| -8092.85[a] | Mg | 0.0000 | 0.0000 | 0.0000 |
| | Mg | 1.5026 | -1.5026 | -2.1240 |
| | Mg | 0.0000 | 0.0000 | -4.2480 |
| | Mg | 3.0052 | 0.0000 | 0.0000 |
| | Mg | 1.5026 | 1.5026 | -2.1240 |
| | Mg | 3.0052 | 0.0000 | -4.2480 |
| | O | 1.5026 | -1.5026 | 0.0020 |
| | O | 0.0000 | 0.0000 | -2.1240 |
| | O | 1.5026 | -1.5026 | -4.2500 |
| | O | 1.5026 | 1.5026 | 0.0020 |
| | O | 3.0052 | 0.0000 | -2.1240 |
| | O | 1.5026 | 1.5026 | -4.2500 |
| | C | 0.0000 | 0.0000 | 2.6100 |
| | O | 0.0000 | 0.0000 | 3.7370 |
| | Vector 1 | 3.0052 | -3.0052 | |
| | Vector 2 | 3.0052 | 3.0052 | |

| | | θ = 0.25 | | |
|---|---|---|---|---|
| Formation energy | Atom | x | y | z |
| -14829.91[b] | Mg | 0.0000 | 0.0000 | 0.0000 |
| | Mg | 1.5026 | 4.5078 | -2.1250 |
| | Mg | 0.0000 | 0.0000 | -4.2500 |
| | Mg | 0.0000 | 3.0052 | 0.0000 |
| | Mg | 1.5026 | 1.5026 | -2.1250 |
| | Mg | 0.0000 | 3.0052 | -4.2500 |
| | Mg | 3.0052 | 0.0000 | 0.0000 |
| | Mg | 4.5078 | 4.5078 | -2.1250 |
| | Mg | 3.0052 | 0.0000 | -4.2500 |
| | Mg | 3.0052 | 3.0052 | 0.0000 |
| | Mg | 4.5078 | 1.5026 | -2.1250 |
| | Mg | 3.0052 | 3.0052 | -4.2500 |
| | O | 1.5026 | 4.5078 | 0.0020 |
| | O | 0.0000 | 0.0000 | -2.1250 |
| | O | 1.5026 | 4.5078 | -4.2500 |
| | O | 1.5026 | 1.5026 | 0.0020 |
| | O | 0.0000 | 3.0052 | -2.1250 |
| | O | 1.5026 | 1.5026 | -4.2500 |
| | O | 4.5078 | 4.5078 | 0.0020 |
| | O | 3.0052 | 0.0000 | -2.1250 |
| | O | 4.5078 | 4.5078 | -4.2500 |
| | O | 4.5078 | 1.5026 | 0.0020 |
| | O | 3.0052 | 3.0052 | -2.1250 |
| | O | 4.5078 | 1.5026 | -4.2500 |
| | C | 0.0000 | 0.0000 | 2.6100 |
| | O | 0.0000 | 0.0000 | 3.7370 |
| | Vector 1 | 6.0104 | 0.0000 | |
| | Vector 2 | 0.0000 | 6.0104 | |

| | | θ = 0.125 | | |
|---|---|---|---|---|
| Formation energy | Atom | x | Y | z |
| -28216.44[c] | Mg | 0.0000 | 0.0000 | 0.0000 |
| | Mg | 1.5026 | -1.5026 | -2.1250 |
| | Mg | 0.0000 | 0.0000 | -4.2500 |
| | Mg | 3.0052 | 0.0000 | 0.0000 |
| | Mg | 1.5026 | 1.5026 | -2.1250 |
| | Mg | 3.0052 | 0.0000 | -4.2500 |
| | Mg | 3.0052 | 3.0052 | 0.0000 |
| | Mg | 4.5078 | 1.5026 | -2.1250 |
| | Mg | 3.0052 | 3.0052 | -4.2500 |

| | | | |
|---|---|---|---|
| Mg | 6.0104 | 3.0052 | 0.0000 |
| Mg | 4.5078 | 4.5078 | -2.1250 |
| Mg | 6.0104 | 3.0052 | -4.2500 |
| Mg | 3.0052 | -3.0052 | 0.0000 |
| Mg | 4.5078 | -4.5078 | -2.1250 |
| Mg | 3.0052 | -3.0052 | -4.2500 |
| Mg | 6.0104 | -3.0052 | 0.0000 |
| Mg | 4.5078 | -1.5026 | -2.1250 |
| Mg | 6.0104 | -3.0052 | -4.2500 |
| Mg | 4.0104 | 0.0000 | 0.0000 |
| Mg | 7.5130 | -1.5026 | -2.1250 |
| Mg | 6.0104 | 0.0000 | -4.2500 |
| Mg | 9.0156 | 0.0000 | 0.0000 |
| Mg | 7.5130 | 1.5026 | -2.1250 |
| Mg | 9.0156 | 0.0000 | -4.2500 |
| O | 1.5026 | -1.5026 | 0.0020 |
| O | 0.0000 | 0.0000 | -2.1250 |
| O | 1.5026 | -1.5026 | -4.2500 |
| O | 1.5026 | 1.5026 | 0.0020 |
| O | 3.0052 | 0.0000 | -2.1250 |
| O | 1.5026 | 1.5026 | -4.2500 |
| O | 4.5078 | 1.5026 | 0.0020 |
| O | 3.0052 | 3.0052 | -2.1250 |
| O | 4.5078 | 1.5026 | -4.2500 |
| O | 4.5078 | 4.5078 | 0.0020 |
| O | 6.0104 | 3.0052 | -2.1250 |
| O | 4.5078 | 4.5078 | -4.2500 |
| O | 4.5078 | -4.5078 | 0.0020 |
| O | 3.0052 | -3.0052 | -2.1250 |
| O | 4.5078 | -4.5078 | -4.2500 |
| O | 4.5078 | -1.5026 | 0.0020 |
| O | 6.0104 | -3.0052 | -2.1250 |
| O | 4.5078 | -1.5026 | -4.2500 |
| O | 7.5130 | -1.5026 | 0.0020 |
| O | 6.0104 | 0.0000 | -2.1250 |
| O | 7.5130 | -1.5026 | -4.2500 |
| O | 7.5130 | 1.5026 | 0.0020 |
| O | 9.0156 | 0.0000 | -2.1250 |
| O | 7.5130 | 1.5026 | -4.2500 |
| C | 0.0000 | 0.0000 | 2.6100 |
| O | 0.0000 | 0.0000 | 3.7370 |
| | | | |
| Vector 1 | 6.0104 | -6.0104 | |
| Vector 2 | 6.0104 | 6.0104 | |

[a] k-space 5, TZ2P(fc), accuracy 5.
[b] k-space 4, TZ2P(fc), accuracy 5.
[c] k-space 3, TZ2P(fc), accuracy 5.

**Table S14.** Cartesian coordinates (in Å) and formation energies for CO on Si(001).[a]

| Formation energy | Atom | x | y | z | Atom | x | y | z |
|---|---|---|---|---|---|---|---|---|
| -30454.45 | Si | 1.9154 | 0.0000 | 1.3544 | Si | 5.6519 | 0.1304 | 6.8209 |
| | Si | 1.9154 | 3.8308 | 1.3544 | Si | 5.6519 | 3.7004 | 6.8209 |
| | Si | 5.7463 | 0.0000 | 1.3544 | Si | 9.6514 | 0.1214 | 6.8130 |
| | Si | 5.7463 | 3.8308 | 1.3544 | Si | 9.6514 | 3.7095 | 6.8130 |
| | Si | 9.5771 | 0.0000 | 1.3544 | Si | 13.3223 | 7.5394 | 6.8192 |
| | Si | 9.5771 | 3.8308 | 1.3544 | Si | 2.0203 | 7.5313 | 6.7744 |
| | Si | 13.4079 | 0.0000 | 1.3544 | Si | 13.3223 | 3.9531 | 6.8192 |
| | Si | 13.4079 | 3.8308 | 1.3544 | Si | 10.5230 | 5.7463 | 7.5395 |
| | Si | 1.9154 | 1.9154 | 2.7088 | Si | 2.8022 | 1.9154 | 7.7299 |
| | Si | 1.9154 | 5.7463 | 2.7088 | Si | 4.7850 | 5.7463 | 7.5116 |
| | Si | 5.7463 | 1.9154 | 2.7088 | Si | 12.4479 | 1.9154 | 7.5404 |
| | Si | 5.7463 | 5.7463 | 2.7088 | Si | 2.5556 | 5.7463 | 8.2559 |
| | Si | 9.5771 | 1.9154 | 2.7088 | Si | 5.1625 | 1.9154 | 8.3614 |
| | Si | 9.5771 | 5.7463 | 2.7088 | Si | 10.2099 | 1.9154 | 8.2908 |
| | Si | 13.4079 | 1.9154 | 2.7088 | Si | 12.7610 | 5.7463 | 8.2979 |
| | Si | 13.4079 | 5.7463 | 2.7088 | H | 0.7801 | 0.0133 | 0.4110 |
| | Si | 11.5014 | 5.7463 | 3.9995 | H | 0.7801 | 3.8176 | 0.4110 |
| | Si | 3.8370 | 1.9154 | 3.9963 | H | 6.8815 | 7.6484 | 0.4110 |
| | Si | 11.4832 | 1.9154 | 3.9997 | H | 6.8815 | 3.8441 | 0.4110 |
| | Si | 3.8273 | 5.7463 | 4.0011 | H | 8.4418 | 7.6484 | 0.4110 |
| | Si | 7.6605 | 1.9154 | 4.1134 | H | 8.4418 | 3.8441 | 0.4110 |
| | Si | 0.0024 | 5.7463 | 4.1176 | H | 14.5432 | 0.0133 | 0.4110 |
| | Si | 15.3214 | 1.9154 | 4.1401 | H | 14.5432 | 3.8176 | 0.4110 |
| | Si | 7.6593 | 5.7463 | 4.1416 | H | 4.5922 | 3.8409 | 0.4378 |
| | Si | 3.8403 | -0.0006 | 5.3049 | H | 10.7311 | 7.6516 | 0.4378 |
| | Si | 3.8403 | 3.8314 | 5.3049 | H | 10.7311 | 3.8409 | 0.4378 |
| | Si | 11.4897 | 7.6615 | 5.3161 | H | 12.2539 | 0.0101 | 0.4378 |
| | Si | 11.4897 | 3.8310 | 5.3161 | H | 12.2539 | 3.8208 | 0.4378 |
| | Si | 15.3202 | 3.8399 | 5.5364 | H | 3.0695 | 0.0101 | 0.4378 |
| | Si | 15.3202 | 7.6526 | 5.5364 | H | 3.0695 | 3.8208 | 0.4378 |
| | Si | 7.6508 | 0.0144 | 5.5315 | H | 4.5922 | 7.6516 | 0.4378 |
| | Si | 7.6508 | 3.8164 | 5.5315 | C | 1.9313 | 1.9154 | 9.3623 |
| | Si | 2.0203 | 3.9612 | 6.7744 | O | 1.3476 | 1.9154 | 10.3618 |
| | Vector 1 | 15.3233 | 0.0000 | | | | | |
| | Vector 2 | 0.0000 | 7.6617 | | | | | |

[a] Energy in kJ mol$^{-1}$. PBE, accuracy 6, k-space 4, TZ2P(fc).

**Table S 15.** Cartesian coordinates (in Å) and formation energies for CO on a $Si_{15}H_{16}$ cluster.[a]

| Formation energy | Atom | X | y | z |
|---|---|---|---|---|
| -14055.60 | Si | 5.1388 | 3.9906 | 2.9892 |
| | Si | 1.5072 | 2.2056 | 4.5297 |
| | Si | 4.6495 | 2.2056 | 4.5297 |
| | Si | 5.1388 | 0.4206 | 2.9892 |
| | Si | 1.5072 | 0.1598 | 2.9427 |
| | Si | 3.3272 | 4.1216 | 1.4732 |
| | Si | 3.3240 | 2.2056 | 0.1647 |
| | Si | 3.3142 | -1.6252 | 0.1695 |
| | Si | 4.2719 | -1.6252 | 3.6800 |
| | Si | 2.2891 | 2.2056 | 3.8983 |
| | Si | 5.1388 | -3.6711 | 2.9892 |
| | Si | 2.0425 | -1.6252 | 4.4243 |
| | Si | 3.3272 | 0.2897 | 1.4732 |
| | Si | 1.5072 | -3.4102 | 2.9427 |
| | Si | 3.3272 | -3.5401 | 1.4732 |
| | H | 2.0880 | -1.6252 | -0.6594 |
| | H | 4.5510 | 2.2056 | -0.6628 |
| | H | 4.5418 | -1.6252 | -0.6572 |
| | H | 2.0944 | 2.2056 | -0.6591 |
| | H | 3.3189 | 5.3450 | 0.6403 |
| | H | 0.2464 | 4.1758 | 2.1713 |
| | H | 4.8381 | -4.7681 | 3.9360 |
| | H | 1.9915 | -4.6773 | 3.5346 |
| | H | 4.5874 | 5.2919 | 3.4286 |
| | H | 3.3251 | -4.7622 | 0.6385 |
| | H | 0.2464 | -3.4858 | 2.1713 |
| | H | 0.2464 | 0.2354 | 2.1713 |
| | H | 6.3810 | 0.3486 | 2.1879 |
| | H | 6.3811 | -3.5990 | 2.1879 |
| | H | 1.8400 | 5.3611 | 3.8637 |
| | H | 6.3811 | 4.0627 | 2.1879 |
| | C | 1.4183 | 2.2056 | 5.5306 |
| | O | 0.8345 | 2.2056 | 6.5302 |

[a] Energy in kJ mol$^{-1}$. PBE, accuracy 6, TZ2P, ADF.

**Table S 16.** Cartesian coordinates (in Å) and formation energy for $C_2H_2$ on Si(001).[a]

| Formation energy | Atom | x | y | z | Atom | x | y | z |
|---|---|---|---|---|---|---|---|---|
| -30505.83 | Si | 1.9154 | 0.0000 | 1.3544 | Si | 5.6801 | 3.8075 | 6.8019 |
| | Si | 1.9154 | 3.8308 | 1.3544 | Si | 9.6669 | 0.1251 | 6.8267 |
| | Si | 5.7463 | 0.0000 | 1.3544 | Si | 9.6669 | 3.7057 | 6.8267 |
| | Si | 5.7463 | 3.8308 | 1.3544 | Si | 13.3181 | 7.5380 | 6.8229 |
| | Si | 9.5771 | 0.0000 | 1.3544 | Si | 1.9775 | 7.5704 | 6.8373 |
| | Si | 9.5771 | 3.8308 | 1.3544 | Si | 13.3181 | 3.9546 | 6.8229 |
| | Si | 13.4079 | 0.0000 | 1.3544 | Si | 10.5342 | 5.7463 | 7.5727 |
| | Si | 13.4079 | 3.8308 | 1.3544 | Si | 2.5396 | 1.9154 | 7.9066 |
| | Si | 1.9154 | 1.9154 | 2.7088 | Si | 4.6711 | 5.7463 | 7.6115 |
| | Si | 1.9154 | 5.7463 | 2.7088 | Si | 12.4478 | 1.9154 | 7.5726 |
| | Si | 5.7463 | 1.9154 | 2.7088 | Si | 2.4369 | 5.7463 | 8.3107 |
| | Si | 5.7463 | 5.7463 | 2.7088 | Si | 4.9041 | 1.9154 | 7.9281 |
| | Si | 9.5771 | 1.9154 | 2.7088 | Si | 10.1900 | 1.9154 | 8.3212 |
| | Si | 9.5771 | 5.7463 | 2.7088 | Si | 12.7829 | 5.7463 | 8.3094 |
| | Si | 13.4079 | 1.9154 | 2.7088 | H | 0.7801 | 0.0133 | 0.4110 |
| | Si | 13.4079 | 5.7463 | 2.7088 | H | 0.7801 | 3.8176 | 0.4110 |
| | Si | 11.5012 | 5.7463 | 4.0001 | H | 6.8815 | 7.6484 | 0.4110 |
| | Si | 3.8368 | 1.9154 | 4.0141 | H | 6.8815 | 3.8441 | 0.4110 |
| | Si | 11.4841 | 1.9154 | 3.9970 | H | 8.4418 | 7.6484 | 0.4110 |
| | Si | 3.8293 | 5.7463 | 4.0172 | H | 8.4418 | 3.8441 | 0.4110 |
| | Si | 7.6569 | 1.9154 | 4.1214 | H | 14.5432 | 0.0133 | 0.4110 |
| | Si | -0.0063 | 5.7463 | 4.1140 | H | 14.5432 | 3.8176 | 0.4110 |
| | Si | -0.0036 | 1.9154 | 4.1535 | H | 4.5922 | 3.8409 | 0.4378 |
| | Si | 7.6686 | 5.7463 | 4.1266 | H | 10.7311 | 7.6516 | 0.4378 |
| | Si | 3.7855 | 0.0014 | 5.3352 | H | 10.7311 | 3.8409 | 0.4378 |
| | Si | 3.7855 | 3.8295 | 5.3352 | H | 12.2539 | 0.0101 | 0.4378 |
| | Si | 11.4904 | 0.0016 | 5.3135 | H | 12.2539 | 3.8208 | 0.4378 |
| | Si | 11.4904 | 3.8292 | 5.3135 | H | 3.0695 | 0.0101 | 0.4378 |
| | Si | -0.0062 | 3.8554 | 5.5414 | H | 3.0695 | 3.8208 | 0.4378 |
| | Si | -0.0062 | 7.6371 | 5.5414 | H | 4.5922 | 7.6516 | 0.4378 |
| | Si | 7.6811 | -0.0011 | 5.5292 | C | 3.0616 | 1.9154 | 9.7465 |
| | Si | 7.6811 | 3.8320 | 5.5292 | C | 4.4197 | 1.9154 | 9.7611 |
| | Si | 1.9775 | 3.9221 | 6.8373 | H | 2.4457 | 1.9154 | 10.6567 |
| | Si | 5.6801 | 0.0233 | 6.8019 | H | 5.0288 | 1.9154 | 10.6751 |
| | Vector 1 | 15.3233 | 0.0000 | | | | | |
| | Vector 2 | 0.0000 | 7.6617 | | | | | |

[a] Energy in kJ mol[-1]. PBE TZ2P(fc), accuracy 5, kspace 1.

**Table S 17.** Cartesian coordinates (in Å) and formation energies for $C_2H_2$ on a $Si_{15}H_{16}$ cluster.[a]

| Formation energy | Atom | x | y | z |
|---|---|---|---|---|
| -14054.47 | Si | 1.9775 | -3.7395 | 6.8373 |
| | Si | 2.5396 | 1.9154 | 7.9066 |
| | Si | 3.8293 | -1.9153 | 4.0172 |
| | Si | 4.9041 | 1.9154 | 7.9281 |
| | Si | 4.6711 | -1.9153 | 7.6115 |
| | Si | 2.4369 | -1.9153 | 8.3107 |
| | Si | 3.8368 | 1.9154 | 4.0141 |
| | Si | 3.7855 | -3.8321 | 5.3352 |
| | Si | 5.6801 | -3.8541 | 6.8019 |
| | Si | 3.7855 | 0.0014 | 5.3352 |
| | Si | 3.7855 | 3.8295 | 5.3352 |
| | Si | 1.9775 | -0.0912 | 6.8373 |
| | Si | 1.9775 | 3.9221 | 6.8373 |
| | Si | 5.6801 | 0.0233 | 6.8019 |
| | Si | 5.6801 | 3.8075 | 6.8019 |
| | H | 6.9288 | -3.8389 | 6.0077 |
| | H | 5.0394 | 5.0386 | 7.3160 |
| | H | 3.8181 | -5.0499 | 4.4947 |
| | H | 5.1882 | -5.0536 | 7.5158 |
| | H | 2.6126 | 1.9154 | 3.1824 |
| | H | 5.0517 | -1.9153 | 3.1829 |
| | H | 5.0586 | 1.9154 | 3.1789 |
| | H | 2.3327 | -5.0075 | 7.5130 |
| | H | 2.2620 | 5.0520 | 7.7499 |
| | H | 2.6075 | -1.9153 | 3.1820 |
| | H | 3.8134 | 5.0488 | 4.4968 |
| | H | 0.7390 | -3.7812 | 6.0282 |
| | H | 0.7390 | -0.0496 | 6.0282 |
| | H | 0.7390 | 3.8805 | 6.0282 |
| | H | 6.9288 | 0.0081 | 6.0077 |
| | H | 6.9288 | 3.8228 | 6.0077 |
| | C | 4.4197 | 1.9154 | 9.7611 |
| | C | 3.0616 | 1.9154 | 9.7465 |
| | H | 5.0288 | 1.9154 | 10.6751 |
| | H | 2.4457 | 1.9154 | 10.6567 |

[a] Energy in kJ mol$^{-1}$. PBE, accuracy 5, TZ2P(fc), ADF.